\documentclass[amstex]{article}
\usepackage{graphicx}
\usepackage{dcolumn}
\usepackage{amsmath}
\usepackage{amssymb}
\usepackage{amsfonts}
\usepackage{amscd}
\begin{document}
\newtheorem{thm}{Theorem}[section]
\newtheorem{lem}[thm]{Lemma}
\newtheorem{prop}[thm]{Proposition}
\newtheorem{cor}[thm]{Corollary}
\newtheorem{assum}{Assumption}[section]
\newtheorem{rem}[thm]{Remark}
\newtheorem{defn}[thm]{Definition}
\newtheorem{rthm}{Theorem}
%%%%%%%%%%%%%%%%%%%%%%%%%%%%%%%%%%%%%%%%%%%%%%%%%%%%%%%%%%%%555
\begin{center}
{\Huge\bf
The Stability of\\} 
\bigskip\bigskip
{\Huge\bf \;the Non-Equilibrium\\}
\bigskip\bigskip
{\Huge\bf Steady States}
\\
\bigskip\bigskip
\bigskip\bigskip
\bigskip\bigskip
{\Large Yoshiko Ogata}
\\
\bigskip
{\it Department of Physics,
Graduate School of Science
\\
The University of Tokyo
\\
7-3-1 Hongo, Bunkyo-ku, Tokyo 
\\
113-0033, Japan}
\end{center}
\vfil
%abstruct
\noindent
We show that the non-equilibrium steady state (NESS) of the free
lattice Fermion model far from equilibrium is macroscopically unstable. 
The problem is translated to that of the spectral analysis
of {\it Liouville Operator}.
We use the method of positive commutators
to investigate it.
We construct a positive commutator on the lattice Fermion system,
whose dispersion relation is $\omega(k)=\cos k-\gamma$.
\noindent
\bigskip
\hrule
\bigskip
\noindent
{\small\tt e-mail: 
ogata@monet.phys.s.u-tokyo.ac.jp}
%%%%%%%%%%%%%%%%%%%%%%%%%%%%%%%%%%
\section{Introduction}
The investigation of equilibrium states in statistical physics has a long history. In mathematical framework, an equilibrium state is defined as a state which satisfies the Kubo-Martin-Schwinger condition. This is a generalization of the Gibbs equilibrium state. Many researches to justify this definition have been made\cite{Haag74},\cite{puszworonowicz} \cite{ruelle1967}.
One of them is the investigation of "{\it return to equilibrium}"; an arbitrary initial state that is {\it normal} with respect to the equilibrium state will converge to the equilibrium state. 
Recall that the notion of {\it normality} represents the macroscopic equivalence
in quasi-local system.
Although it is a physically fundamental phenomenon, to prove it rigorously is not easy. A complete proof of return to equilibrium in the two-sided XY-model was given by H. Araki \cite{A84}. The other example is an open quantum system, which consists of a finite subsystem and an infinitely extended reservoir in equilibrium \cite{JPf2},\cite{JPf3},\cite{Merk01}.
The open quantum system converges to the asymptotic state, which is the equilibrium state of the coupled systems.
The notable fact here is that this state
is normal to the initial one, i.e., macroscopically equivalent to the initial
state. 
This shows that the equilibrium state of the reservoir is macroscopically stable when a finite subsystem is connected to.

Recently non-equilibrium steady state (NESS) far from equilibrium has attracted considerable interests. The NESS is introduced as a state asymptotically realized from an inhomogeneous initial state \cite{JPNESS1},\cite{JPNESS2},
\cite{ruelle00}. A question rises naturally here; is the NESS macroscopically stable? As an analogy of return to equilibrium in open system, we connect a finite small system to the NESS through a bounded interaction.
Will the NESS 
converge to a state that is normal to the initial state or not?
We consider this problem about
a free Fermion model on one-dimensional lattice. 
The explicit form of the NESS is known on this model
\cite{HA00},\cite{Pillet}.
We show if the NESS is far from equilibrium, it is macroscopically unstable.
This results is due to the following fact:
for the NESS far from equilibrium,
the number of the particles with momentum $k$
is different from the number of the particles with momentum $-k$,
although they have the same energy.

 Technically, the investigation corresponds to the study of the spectral property about the Liouville operator, which represents the dynamics on the 
 Gelfand Naimark Segal (GNS) Hilbert space of the initial state.  We use the positive commutator method to analyze the spectrum. It is well-known that a radiation field system with energy dispersion relation 
$\omega(k)=\vert k\vert$ has a nice covariance property
and the positive commutator is constructed with the aid of it. However, in our system, the dispersion relation is $\omega(k)=\cos k$, a new method for constructing a positive commutator is required. We shall construct such a positive commutator and investigate the macroscopic stability of the NESS in this paper. 

The paper is organized as follows. In the next section, we introduce basic definitions and notations, then state the main theorems. 
In Section \ref{strategy}, we will explain the strategy of the proof.
In Section \ref{stand}, we review the role of
the standard theory in the research of the NESS, and introduce
the modular structure of our model.
In Section \ref{rescale}, we introduce the rescaling group which is the key
to construct the positive commutator.
In order for the positive commutator method to work,
we have to introduce the cut off of the interaction.
This is done in Section \ref{cut}.
Section \ref{PC} is devoted for the construction of positive commutator.
Then in Section \ref{Virial}, we derive the spectral property of the
Liouville operator using the method of M.Merkli \cite{Merk01}
on Virial Theorem.
We complete the proof in Section \ref{stability}.
Below, for a self-adjoint operator $A$, we denote by $P(A\subset I)$
the spectral projection of $A$ onto the subset $I$.
%%%%%%%%%%%%%%%%%%%%%%%%%%%%%%%%%%%%%%%%%%%%%%%%%%
\section{Main Results}
In this section, we introduce the basic definitions, and state
the main results.
\subsection{\it The $C^*$-algebraic Framework}
A $C^*$-dynamical system is a pair $(\cal{O},\tau)$
where $\cal O$ is a $C^*$-algebra, and $\tau$ is a
strongly continuous one-parameter group of automorphisms of $\cal O$.
The elements of $\cal O$ describe observables in the physical system
and $\tau$ specifies their time evolution.
Below we assume that $\cal O$ has an identity.
A physical state is described as a positive linear functional 
with norm $1$.
Let $\omega$ be a state on $\cal O$ with GNS triple
$({\cal H},\pi,\Omega)$.
The notion of $\omega$-normal is defined as follows:
\begin{defn}\label{normal}
A state $\eta$ is said to be $\omega$-normal
if there exists a density matrix $\rho$ on ${\cal H}$
such that $\eta(\cdot)={\rm Tr}\rho\pi(\cdot)$.
\end{defn}
If a state is not $\omega$-normal, it is called $\omega$-singular.
For quasi-local algebra, $\omega$-normality means that 
$\eta$ is approximated in norm topology by
local perturbation of $\omega$.
So in quasi-local algebra, we can make the following interpretation:
if $\eta$ is $\omega$-normal and $\omega$ is $\eta$-normal,
$\omega$ and $\eta$ are macroscopically equivalent. 

The NESS of dynamics $\tau_t$
associated with the state
$\omega$ are the weak$-*$ accumulation points of the set of states
\begin{align*}
\frac{1}{T}\int_0^T\omega\circ\tau_t dt
\end{align*}
as $T\to\infty$.
We denote the set of the NESS by $\Sigma_{\tau}(\omega)$. 
As the set of states on $\cal O$
is weak $*$ -compact by Alaoglu's Theorem, 
$\Sigma_\tau(\omega)$ is a 
non-empty set whose elements are $\tau$-invariant. 

\subsection{\it Macroscopic instability}
In this subsection, we introduce the notion of {\it macroscopic instability}.
Let $({\cal O}_m,\tau_m)$ be a $C^*$-dynamical system.
Let $\omega_0$ be an initial state over ${\cal O}_m$ and let
$\omega_m$ be the NESS corresponding to the pair 
$(\omega_0,\tau_m)$ i.e.,
$\omega_m\in{\Sigma_{\tau_m}}(\omega_0)$.
Now we are interested in the stability of $\omega_m$.
To investigate it, we add an external finite $C^*$-dynamical system
$({\cal O}_S,\tau_S)$.
The combined system $({\cal O}_S\otimes{\cal O}_m,\tau_S\otimes\tau_m)$
is also a $C^*$-dynamical system.
Let us introduce a bounded interaction $V$ between ${\cal O}_S$ and
${\cal O}_m$,
and denote by $\tau_V$ the perturbed dynamics.
We shall define the {\it macroscopic instability} of $\omega_m$ as follows:
\begin{defn}
The NESS 
$\omega_m\in\Sigma_{\tau_m}(\omega_0)$ is {\it macroscopically unstable
under a perturbation $V$},
if for any state $\omega_S$ over ${\cal O}_S$,
no element in $\Sigma_{\tau_V}(\omega_S\otimes\omega_m)$ 
is  $\omega_S\otimes\omega_m$-normal.
The NESS $\omega_m\in\Sigma_{\tau_m}(\omega_0)$ is 
{\it macroscopically unstable} if $\omega_m$
is macroscopically unstable under some bounded perturbation $V$.
\end{defn}
Let us explain it in detail for our system.
First we divide the one-dimensional Fermion lattice to the left and the right,
and consider a state that each side is in equilibrium
at different temperature. 
This is the initial state $\omega_0$.
The corresponding NESS $\Sigma_{\tau_m}(\omega_0)$
under the time evolution of free lattice Fermion ${\tau_m=\alpha^f}$ 
consists of  
only one point: $\Sigma_{\tau_m}(\omega_0)=\{\omega_\rho\}$. 
To investigate the stability of $\omega_m=\omega_\rho$,
we prepare an external finite system 
described in a finite dimensional Hilbert space ${\mathbb C}^d$.
We connect it to $\omega_m$
through a bounded interaction $V$.
If there is no $\omega_S\otimes\omega_m$-normal state in 
$\Sigma_{\tau_V}(\omega_S\otimes\omega_m)$, 
for any state $\omega_S$ over ${\cal O}_S$,
$\omega_m$ is macroscopically unstable.

\subsection{\it The model}
In this paper we consider the free lattice Fermion system
in one dimension.
The explicit form of the NESS is known for this system \cite{HA00},\cite{Pillet}.

Let ${\mathfrak h}\equiv l^2({\mathbb Z})$ be a Hilbert space of
a single Fermion.
By Fourier transformation,
it is unitary equivalent to $L^2([-\pi,\pi))$.
The Hamiltonian  $h$ of a single Fermion is given by
\begin{align}
\left(h f\right)\left(n\right)
=\frac 12 \left(f\left(n-1\right)+f\left(n+1\right)\right)
-\gamma f\left(n\right),
\label{onehamiltonian}
\end{align} 
on $\mathfrak h$
which is described in Fourier representation as
\begin{align*}
\widehat{hf}\left(k\right)
&=\omega\left(k\right){\hat f}\left(k\right),\\
\omega\left(k\right)&=\cos\left(k\right)-\gamma.
\end{align*} 
The $\gamma$-term represents the interaction with the external field,
and $\gamma$ is a parameter in $(-1,1)$. 
The free lattice Fermi gas ${\cal R}$ is 
described as the CAR-algebra ${\cal O}_f$
over ${\mathfrak h}$.
And its dynamics is given by 
\begin{align}
\alpha_t^f\left( a\left( f \right)\right)
=a\left({\rm e}^{ith} f \right).
\label{tau}
\end{align}

In the initial state $\omega_0$,
the lattice is separated into the left and the right.
And they are kept at different inverse temperature
$\beta_-,\beta_+$, respectively.
 The NESS $\omega_\rho$ associated with $\omega_0$
 is realized as the asymptotic state
 under the dynamics $\alpha^f$ (\ref{tau}). 
The explicit form of $\omega_\rho$ was obtained in
\cite{HA00} and \cite{Pillet}, independently:
$\omega_\rho$ is a state whose $n$-point functions have a structure
\begin{align}
\omega_\rho
\left(a(f_n)^*\cdots a(f_1)^*a(g_1)\cdots a(g_m)\right)
=\delta_{nm}\det (\langle f_i,\rho g_j\rangle),
\label{Fdis}
\end{align}
where $\rho$ is represented as a multiplication operator,
\begin{align}
\rho(k)&=\left\{
\begin{gathered}
\left(1+e^{\beta_{+}\omega(k)}\right)^{-1}
\quad k\in[0,\pi)\\
\left(1+e^{\beta_{-}\omega(k)}\right)^{-1}
\quad k\in[-\pi,0)
\end{gathered},
\right. \label{rho}\\
\omega(k)&=\cos\left(k \right)-\gamma,\nonumber
\end{align}
in the Fourier representation.
If $\beta_+\neq\beta_-$,
we will say that the NESS $\omega_\rho$ is far from equilibrium.
In this paper, the stability of this state 
is considered.

The observables of the small system
are described as $C^*$-algebra ${\cal O}_S\equiv B({\mathfrak H}_S)$
on a finite $d$-dimensional Hilbert space
${\mathfrak H}_S={\mathbb C}^d$.
We denote by $H_S$ the free Hamiltonian of the system
on ${\mathfrak H}_S$.
The free dynamics $\alpha_t^S$ is given by
\begin{align*}
\alpha_t^S\left(A\right)={\rm e}^{itH_S}A{\rm e}^{-itH_S}.
\end{align*} 
The combined system $\cal S+R$ is described as the $C^*$-algebra
\begin{align*}
{\cal O}\equiv {\cal O}_S\otimes{\cal O}_f.
\end{align*} 
The free dynamics of the combined system is given by
$\alpha_t^0=\alpha_t^S\otimes\alpha_t^f$.
We denote by $\delta_0$ the derivation of $\alpha_t^0$.

Let us consider the dynamics including the interaction between
$\cal S$ and $\cal R$.
In this paper, we define the interaction term $V$ by
\begin{align}
V
=\lambda\cdot Y\otimes\left(a\left(f\right)+a^*\left(f\right)\right),
\label{V}
\end{align} 
where $f\in{\mathfrak h}$ is called a form factor.
Here $\lambda$ is a coupling constant and 
$Y$ is a self-adjoint operator on ${\mathfrak H}_S$.
Note that $V$ is an element of $\cal O$.
The perturbed dynamics $\alpha_t$ 
is generated by $\delta=\delta_0+i[V,\cdot]$
with $D(\delta)=D(\delta_0)$.
$\alpha_t$ is expanded as follows;
\begin{align}
\alpha_t\left(A\right)
\equiv\alpha_t^0\left(A\right)
+\sum_{n\ge 1}i^n
\int_0^tdt_1\int_0^{t_1}dt_2\cdots\int_0^{t_{n-1}}dt_n
\left[\alpha_{t_n}^0\left(V\right),
\left[\cdots,\left[\alpha_{t_1}^0\left(V\right),
\alpha_t^0\left(A\right)\right]\right]\right].
\label{cper}
\end{align}
The right hand side converges in norm topology in $\cal O$.
$\alpha_t$ is strongly continuous one parameter group of
automorphisms.
%%%%%%%%%%%%%%%%%%%%%%%%%%%%%%%%%%%%%%%%%%%%%
%%%%%%%%%%%%%%%%%%%%%%%%%%%%%%%%%%%%%%%%%%%%%%%%%%
\subsection{\it Main Theorem}
In the analysis, 
we carry out the variable transformation from $k\in [-\pi,\pi)$
to $t\in\mathbb R$, by
$t(k)=\tan\frac{k}{2}$.
Under this variable translation, we identify $\mathfrak h=L^2([-\pi,\pi),dk)$
with $L^2({\mathbb R},\frac{2}{t^2+1}dt)$.
We need several assumptions
on the small system and the interaction $V$.
\begin{assum}\label{suppg}
Under the identification 
$\mathfrak h=L^2([-\pi,\pi),dk)=L^2({\mathbb R},\frac{2}{t^2+1}dt)$,
let $u(\theta)$ be 
a strongly continuous one parameter unitary group
on $\mathfrak h$ defined by 
\begin{align*}
\left(u(\theta) g\right)\left(t\right)
\equiv{\rm e}^{\frac{\theta}{2}}
\sqrt{\frac{t^2+1}{{\rm e}^{2\theta}t^2+1}}
g\left({\rm e}^{\theta}t\right),
\quad \theta\in{\mathbb R},\quad
g\in L^2((-\infty,\infty),\frac{2}{t^2+1}dt),
\end{align*}
and let $p$ be the generator of $u(\theta)={\rm e}^{i\theta p}$.
For a constant $0<v<1$, let 
$\Lambda_v$ be the interval of $\mathbb R$ defined by
\[\Lambda_v=\{t\in{\mathbb R}\;;\;\;s_1(t)=\frac{4t^2}{(1+t^2)^2}\ge v\}.\]
We assume that the form factor $f\in{\mathfrak h}$ in (\ref{V})
is in the domain $D(p^3)$.
Furthermore, we assume that 
there exists $0<v<1$ such that the support of $f$ satisfies
\[{\rm supp}f\subset \Lambda_v.\]
\label{assuminter}
\end{assum}
\begin{assum}\label{O}
Let $e_i,e_j$ be an arbitrary pair of eigenvalues of 
$L_S=H_S\otimes 1-1\otimes H_S$
on ${\mathfrak H}_S\otimes{\mathfrak H}_S$
such that $e_i\neq e_j$.There exists $k_{ij}$ 
such that
\begin{align*}
\omega(\pm k_{ij})=\cos k_{ij}-\gamma=e_i-e_j,\\
0\le k_{ij},\quad t(k_{ij})\in{\Lambda_v}.
\end{align*}
\end{assum}
\begin{assum}\label{nondegenerate}
\begin{enumerate}
\item
The Hamiltonian $H_S$ has no degenerated eigenvalue.
\item
There exists $a>0$ such that
\[
\min_{e_i\neq e_j}\vert f(\pm k_{ij})\vert
\ge a.
\]
\end{enumerate}
\end{assum}
\begin{assum}\label{small2}
Let $\varphi_n$ be the $n$-th eigenvector of $H_S$
with eigenvalue $E_n$, and let $p_n$ be the spectral
projection onto $\varphi_n$ on ${\mathfrak H}_S$.
We denote the eigenvalue of 
$L_S=H_S\otimes 1-1\otimes H_S$ as $E_{n,m}\equiv E_n-E_m$,
and introduce the following subsets for each eigenvalue $e$ of $L_S$:
\begin{align*}
N_l^{(j)}\equiv\{i;E_{i,j}=e\},\quad
N_r^{(i)}\equiv\{j;E_{i,j}=e\}\\
N_l\equiv\bigcup_j N_l^{(j)},\quad
N_r\equiv\bigcup_i N_r^{(i)}.
\end{align*}
For a set $\cal N$, we define a projection
$p_{\cal N}\equiv\sum_{n\in{\cal N}}p_n$.
\begin{enumerate}
\item
For $e\neq 0$, we assume
\[2\delta_0\equiv
\min_{n\in N_l}\inf\sigma\left(\left.
p_{N_r^{(n)}}\bar Yp_{N_r^c}\bar Yp_{N_r^{(n)}}
\right\vert_{p_{N_r^{(n)}}{\mathfrak H}_S}\right)
+\min_{m\in N_r}\inf\sigma\left(\left.
p_{N_l^{(m)}}Yp_{N_l^c}Yp_{N_l^{(m)}}
\right\vert_{p_{N_l^{(m)}}{\mathfrak H}_S}
\right)>0,
\]
where $\sigma(A)$ represents the spectrum of $A$,
and $\bar Y$ is the complex conjugation of $Y$ with respect to
the orthonormal basis consisting of eigenvectors of $H_S$.
\item
Let $Y_{mn}\equiv 
\langle\varphi_m\vert Y\vert\varphi_n\rangle$.
We assume
$\vert Y_{mn}\vert>0$ for all $m\neq n$
\end{enumerate}
\end{assum}
An example which satisfies all of the Assumptions is represented in Appendix
\ref{exam}.
Here is the main theorem of this paper.
\begin{thm}\label{unstable}
Let $\omega_\rho$ be a NESS far from equilibrium, i.e., the state given by
(\ref{Fdis}) with inverse temperatures $\beta_+\neq\beta_-$.
Suppose that Assumption \ref{suppg} to \ref{small2} are satisfied.
Then there exists a $\lambda_1>0$ s.t. if
$0<\vert \lambda \vert<\lambda_1$, then $\omega_\rho$ is macroscopically
unstable under the
perturbation $V$ (\ref{V}).
Especially, $\omega_\rho$ is macroscopically unstable.

Furthermore, assume $\beta_0<\beta_+,\beta_-<\beta_1$,
$\Vert p f\Vert,\Vert f\Vert\le b$ for any fixed 
$0<\beta_0<\beta_1<\infty$ and $0<b<\infty$.
Here $p$ is the generator of $u(\theta)={\rm e}^{i\theta p}$.
Then if we fix $\beta_+$ and $\beta_-$, we have
$\lambda_1\sim O(v^{\frac{50}{9}})$ as $v$ goes to $0$.
On the other hand, if we fix $v$ then we have 
$\lambda_1\sim O(\vert\beta_+-\beta_-\vert^{\frac{200}{11}})$
as $\beta_+-\beta_-\to 0$.
\end{thm}
\begin{thm}\label{frte}
Let $\omega_\rho$ be an equilibrium state i.e. the state given in (\ref{Fdis})
with $\beta\equiv \beta_+=\beta_-$.
Then there exists a $\beta-$KMS state $\omega_V$ w.r.t. 
the perturbed dynamics $\alpha_t$ (\ref{cper}), which is normal to 
$\omega_S\otimes\omega_\rho$ for arbitraly faithful state $\omega_S$
of ${\cal O}_S$.
Suppose that Assumption \ref{suppg} to \ref{small2} are satisfied.
Then there exists a $\lambda_1>0$ such that if
$0<\vert\lambda\vert <\lambda_1$, any $\omega_V$-normal state $\eta$
exhibits return to equilibrium in an
ergodic mean sense,i.e.
\[
\lim_{T\to\infty}\frac{1}{T}
\int_0^T\eta\left(\alpha_t(A)\right)dt
=\omega_V(A)
\]
for all $A\in{\cal O
}$.

Furthermore, assume $\beta_0<\beta<\beta_1$,
$\Vert p f\Vert,\Vert f\Vert\le b$ for any fixed 
$0<\beta_0<\beta_1<\infty$ and $0<b<\infty$.
Then we have
$\lambda_1\sim O(v^{\frac{50}{9}})$ as $v$ goes to $0$.
\end{thm}
%%%%%%%%%%%%%%%%%%%%%%%%%%%%%%%%%%%%%%%%%%%%%%%%%%%%%%%%%%%
\section{The Strategy of the Proof}\label{strategy}
In this section, we explain the strategy and the organization of the proof.
\subsection{\it The kernel of Liouville operator}
By the standard theory, the problem of 
macroscopic instability is translated into
the spectral problem of so called Liouville operator $L$
(see Section \ref{stand} Proposition \ref{sigular})
: if ${\rm Ker}L=\{0\}$, $\omega$ is macroscopically unstable.
In our model, the Liouville operator $L$ is an operator
on ${\cal H}=({\mathfrak H}_S\otimes{\mathfrak H}_S)\otimes {\cal F}
({\mathfrak h}\oplus{\mathfrak h})$, given by
\begin{align}
&L=
\left(H_S\otimes 1\right)\otimes 1-
\left(1\otimes H_S\right)\otimes 1 +
1\otimes d\Gamma\left(h\oplus -h\right)
+\lambda I_0.
\label{defL}
\end{align}
Here ${\cal F}({\mathfrak h}\oplus{\mathfrak h})$
is a Fermi Fock space over ${\mathfrak h}\oplus{\mathfrak h}$ and
$d\Gamma\left(h\oplus -h\right)$is a secound quantization of
the multiplication operator $h\oplus -h.$
$\lambda I_0$ is the interaction term.\\
The main part of this paper is to prove
the following Theorem on the eigenvector of $L$:
\begin{thm}\label{spectral}
For each eigenvalue ${\tilde e}$ of $L_S=H_S\otimes 1-1\otimes H_S$,
define an operator $\Gamma({\tilde e})$ on 
the space $P(L_S={\tilde e})\cdot{\mathfrak H}_S\otimes{\mathfrak H}_S$ by
\begin{align}
\Gamma({\tilde e})\equiv
\int_{-\pi}^\pi dk \;m(k,1)^*P\left(L_S\neq {\tilde e}\right)
\delta\left(\omega\left(k\right)+L_S-{\tilde e}\right)m(k,1)\nonumber\\
+\int_{-\pi}^\pi dk \;m(k,2)^*P\left(L_S\neq {\tilde e}\right)
\delta\left(-\omega\left(k\right)+L_S-{\tilde e}\right)m(k,2)\nonumber\\
m(k,i)\equiv Y\otimes 1\cdot g_1^i(k)-1\otimes{\bar Y}\cdot g_2^i(k)
\quad i=1,2.
\label{gammae}
\end{align}
Here $g_i^j$ are defined by
\[
g_1^1=\left(1-\rho\right)^{\frac12}f,\quad
g_1^2=\rho^{\frac12}{\bar f},
\quad\;g_2^1=\rho^{\frac12}f,\quad
g_2^2=\left(1-\rho\right)^{\frac12}{\bar f},
\]
and $\bar f$ is the complex conjugation of $f$ 
in the Fourier representation.
Let $\gamma_{\tilde e}$ be a strictly positive constant such that
\[
\Gamma(\tilde e)\ge \gamma_{\tilde e}
\cdot \left(P(\Gamma(\tilde e)=0)\right)^{\perp}.
\]
Let $\tilde P_{\tilde e}$ be
\[
{\tilde P}_{{\tilde e}}=P(L_S={\tilde e})\cdot P(\Gamma({\tilde e})=0)\otimes P_{\Omega_f},
\]
where $P_{\Omega_f}$ is the projection onto 
the vacuum $\Omega_f$ of ${\cal F}({\mathfrak h}\oplus{\mathfrak h})$.\\
For $e\in{\mathbb R}$, let 
$\tilde e(e)\equiv e$ if $e$ is an eigenvalue of $L_S$,
and let $\tilde e(e)$ be an eigenvalue of $L_S$
which is next to $e$, if $e$ is not an eigenvalue of
$L_S$.
Suppose that Assumption \ref{suppg} and \ref{O} are satisfied.
Then there exists $\lambda_1>0$ s.t.,if $0<\vert\lambda\vert<\lambda_1$, then
there is no eigenvector with eigenvalue $e$
which is orthogonal to $\tilde P_{\tilde e(e)}$.
In particular, if $\tilde P_{\tilde e(e)}=0$, $e$
is not an eigenvalue of $L$.
Furthermore,
if we assume $\beta_0<\beta_+,\beta_-<\beta_1$,
$\Vert p f\Vert,\Vert f\Vert\le b$ for any fixed 
$0<\beta_0<\beta_1<\infty$ and $0<b<\infty$,
we can choose $\lambda_1$ as
\begin{align}
\lambda_1=C\min\{
v^{\frac{100}{26}},\left(\frac{v}{\gamma_{{\tilde e}(e)}}\right)^{\frac{100}{182}},
(\gamma_{{\tilde e}(e)})^{\frac{100}{11}},(v\gamma_{{\tilde e}(e)})^{\frac{100}{18}}
\}\label{lam1}.
\end{align}
Here $C$ is a constant which depends on $\beta_0,\beta_1$ and $b$, but
is independent of $v$ and $\beta_+-\beta_-$.
\end{thm} 
By Theorem \ref{spectral}, the existence of the eigenvector
of $L$ is determined by the kernel of $\Gamma(e)$.
We have the following Theorem on it (Section \ref{stability}):
\begin{thm}\label{gamma}
Suppose that Assumption \ref{suppg} to \ref{small2} are satisfied.
Then if $\beta_+\neq\beta_-$, ${\rm Ker}\Gamma(e)=\{0\}$
for any eigenvalue $e$ of $L_S$.
If $\beta_+=\beta_-$, ${\rm Ker}\Gamma(e)=\{0\}$
for any non-zero eigenvalue $e\neq 0$ of $L_S$,
but ${\rm Ker}\Gamma(0)$ is non-trivial one-dimensional space.
\end{thm}
The following fact will develop in the proof :
the non-existence of the non-trivial kernel of $\Gamma(0)$
for the far from equilibrium case $\beta_+\neq\beta_-$
is caused by the fact that the number of the particles
with the momentum $k,-k$
are different although they have the same energy $\omega(k)=\omega(-k)$.
Note that the same fact induces the existence of current
for NESS.

Combining Theorem \ref{spectral} and \ref{gamma},
we obtain the instability of NESS far from equilibrium.

\subsection{\it The proof of Theorem \ref{spectral}}
To prove the Theorem,
we use the positive commutator method.
The idea of the method is as follows:
suppose that there exists anti-self-adjoint operator $A$
such that
$[L,A]\ge c>0$.
If $L$ has an eigenvector $\psi$ with eigenvalue $e$,
we have 
\begin{align*}
0=2{\rm Re}\left\langle (L-e)\psi,A\psi\right\rangle
=\left\langle\psi, [L-e,A]\psi\right\rangle
\ge c>0,
\end{align*}
which is a contradiction.
Hence $L$ has no eigenvector.
This is called Virial Theorem.
However this argument works only formally
and indeed we have to take care of the domain question.

The proof of Theorem \ref{spectral} is divided into two steps:
the first one is to construct the positive commutator
(Section \ref{rescale} to Section \ref{PC}),
and the second one is to justify the above arguments rigorously
(Section \ref{Virial}).

To construct the positive commutator is a non-trivial problem,
and we have to work out for each models.
Now, our field has the dispersion relation $\omega(k)=\cos k-\gamma$.
Under the variable translation $k\to t$, 
we show that there exists a strongly continuous
one parameter group of unitaries $U(\theta)$ on $\cal H$,
which satisfies
\begin{align*}
U(-\theta)\left(1\otimes d\Gamma\left(h\oplus -h\right)\right)U(\theta)=
1\otimes d\Gamma\left(h^{-\theta}\oplus -h^{\theta}\right),
\end{align*}
where $h^\theta$ is a multiplication operator on $L^2({\mathbb R},\frac{2}{t^2+1}dt)$  defined by
$h^\theta(t)\equiv h({\rm e}^{\theta}t)$.
This means $U(\theta)$ induces the rescaling of multiplication operator
in $\mathfrak h$.
Let $A_0$ be the generator of $U(\theta)$.
This $A_0$ satisfies
\begin{align*}
\left[L_0,A_0\right]=S_1=1\otimes d\Gamma\left(s_1\oplus s_1\right)\ge 0
\end{align*}
where $S_1$ is a second quantization
of multiplication operator 
\begin{align*}
\left(s_1f\right)(t)=s_1(t)f(t),\quad
s_1(t)=\frac{4t^2}{(1+t^2)^2}.
\end{align*}
Hence by considering rescaling of multiplication operators with respect to $t$,
we obtained positive commutator $[L_0,A_0]=S_1\ge 0$.
However, what is really needed is the strictly positive commutator.
If $S_1$ has a spectral gap, following the well-known procedure,
we can construct a strictly positive commutator.
But $S_1$ does not have a gap now, and we need to overcome this problem.

Let ${\Lambda_v^c}$ be the complement of ${\Lambda_v}$,
and let $N_{\Lambda_v^c}$ be the number operator of particles
whose momentum is included in $\Lambda_v^c$.
Furthermore, let $P$ be the spectral projection onto the subspace
$N_{\Lambda_v^c}=0$, and set $\bar P=1-P$.
If we restrict ourselves to $P{\cal H}$,
$PS_1P$ has a spectral gap $v>0$,
so if $L$ strongly commutes with $N_{\Lambda_v^c}$,
we can construct the positive commutator 
for $L$ in $P{\cal H}$.
Furthermore, if the subspace 
$\bar P{\cal H}$ includes no eigenvector of $L$,
we just need to analyze $L$ on $P{\cal H}$.

To make the spectral localization possible,
we introduce a cut off of the form factor $f\in{\mathfrak h}$
with respect to $\Lambda_v$,i.e.,
\begin{align*}
supp f\subset\Lambda_v.
\end{align*}
By this cut off, $L$ strongly commutes with $N_{\Lambda_v^c}$,
and furthermore, there is no eigenvector of $L$,
in the subspace $\bar P{\cal H}$.
Hence we obtain positive commutator which is sufficient to analyze the
eigenvector of $L$.

To carry out the second step, the rigorous justification
of the Virial Theorem, we used the new method introduced by M.Merkli.
By approximating the eigenvector
of $L$ by vectors in the domain of total number operator $N$ and $A_0$,
we can carry out the arguments rigorously. 

%%%%%%%%%%%%%%%%%%%%%%%%%%%%%%%%%%%%%%%%%%%%%%%%%%%%%%%%
\section{The Standard Theory and NESS}{\label {stand}}
In this section, we 
explain the role of the standard theory (see \cite{BR86})
in the investigation of the NESS
and introduce the modular structure in our model.
\subsection{\it The standard theory}
Let $\cal M$ be a von Neumann algebra
on a Hilbert space ${\cal H}$
with a cyclic and separating vector $\Omega$.
A positive linear functional $\omega$ on $\cal M$ is 
said to be {\it normal} if there exists a positive traceclass operator $\rho$ such that
\begin{align*}
\omega\left(\cdot\right)={\rm Tr}\left(\rho\cdot\right).
\end{align*} 
We denote by $\cal M_{*+}$
the set of all normal positive linear functionals. 

We can define an operator $S_0$ on a dense set
${\cal M}\Omega$ by
\begin{align*}
S_0x\Omega=x^*\Omega.
\end{align*}
$S_0$ is closable and we represent the polar decomposition 
of the closure $\bar S_0$ as
\begin{align*}
\bar S_0=J\Delta^{\frac 12}.
\end{align*}
$J$ is called the {\it modular conjugation} and
$\Delta$ is called the {\it modular operator}.
By the Tomita-Takesaki Theory we have $J{\cal M}J={\cal M}'$,
where $\cal M'$ is the commutant of $\cal M$.
Let $j:{\cal M}\to{\cal M}'$ be the antilinear $*$-isomorphism
defined by $j(x)=JxJ$. 
We define the {\it natural positive cone} $\cal P$
by the closure of the set
$\{ xj\left(x\right)\Omega ; x\in{\cal M}\}$.
For $\xi\in{\cal P}$, 
define the normal positive functional
$\omega_\xi\in{\cal M}_{*+}$ by
\begin{align*}
\omega_\xi\left(x\right)=\left\langle\xi,
x\xi\right\rangle
\quad \forall x\in{\cal M}.
\end{align*}
We have the following Theorems;
\begin{thm}
For any $\omega\in\cal M_{*+}$,
there exists a unique $\xi(\omega)\in{\cal P}$ such that
$\omega=\omega_{\xi\left(\omega\right)}$.
\label{stand1}
\end{thm}
\begin{thm}
For any $*$-automorphism $\alpha$ of $\cal M$,
there exists a unique unitary operator $U(\alpha)$ on $\cal H$
satisfying the following properties:
\begin{enumerate}
\item $U\left(\alpha\right)xU\left(\alpha\right)^*
=\alpha\left(x\right),\quad x\in {\cal M};$
\item $U\left(\alpha\right){\cal P} \subset {\cal P}$\;
and $U\left(\alpha\right)\xi\left(\omega\right)
=\xi\left(\alpha^{-1*}\left(\omega\right)\right),
\;\; \omega\in\cal M_{*+}$,\\
where $(\alpha^*\omega)(A)\equiv\omega(\alpha(A))$,
\item $\left[U\left(\alpha\right),J \right]=0$.
\end{enumerate}
\label{stand2}
\end{thm}
Let $\alpha_t$ be a one parameter group of automorphisms and let
$U(t)$ be a one parameter group of unitaries associated with $\alpha_t$.
If $U(t)$ is strongly continuous, it is written as $U(t)={\rm e}^{itL}$
with self adjoint operator $L$.
We call $L$ the Liouville operator of $\alpha_t$.
%%%%%%%%%%%%%%%%%%%%%%%%%%%%
\subsection{\it The NESS and the Liouville operator}
Now we move from $W^*$-dynamical systems
to $C^*$-dynamical systems, and explain the role of the standard theory
in the investigation of the NESS.
Let $\alpha_t$ be a one parameter group of automorphisms
which describes the dynamics of a unital $C^*$-algebra $\cal O$,
and let $\omega$ be a state of $\cal O$.
Let $({\cal H},\pi,\Omega)$ be
the GNS triple of $\omega$.
Suppose that $\Omega$ is a cyclic and separating vector for von Neumann
algebra $\pi(\cal O)^{''}$, and that there exists
an extension $\tilde\alpha_t$ of $\alpha_t$ 
to $\pi(\cal O)^{''}$.
Then by Theorem \ref{stand2} there is one parameter group of
unitary operators $U_t$ such that
\begin{align}
&\tilde\alpha_t\left(x\right)=U_txU_t^*
\quad x\in\pi(\cal O)^{''}\nonumber\\
&U_t^*\cal P\subset\cal P.
\label{up}
\end{align}
In addition, suppose that $U_t$ is strongly continuous
and let $L$ be the Liouville operator.
If $\omega_{\infty}\in\Sigma_\alpha(\omega)$ is $\omega$-normal,
there exists a vector $\xi_\infty$ in
${\cal P}$ s.t.
\begin{align*}
\omega_\infty\left(A\right)=\left\langle\xi_\infty,
\pi\left(A\right)\xi_\infty\right\rangle,
\quad A\in\cal O,
\end{align*}
by Theorem \ref{stand1}.
As $\omega_{\infty}\in\Sigma_\alpha(\omega)$,
$\omega_\infty$ is invariant under $\alpha_t$:
$\omega_{\infty}\circ\alpha_t=\omega_{\infty}$.
Hence we have
\begin{align*}
\left\langle U_t^*\xi_\infty,
\pi\left(A\right)U_t^*\xi_\infty\right\rangle.
=\omega_\infty\circ\alpha_t\left(A\right)
=\omega_\infty\left(A\right)
=\left\langle\xi_\infty,
\pi\left(A\right)\xi_\infty\right\rangle,
\quad A\in\cal O.
\end{align*}
As $\xi_\infty,U_t^*\xi_\infty\in{\cal P}$, we get
\begin{align*}
U_t^*\xi_\infty=\xi_\infty,
\end{align*}
by Theorem \ref{stand1}.
This means
\begin{align*}
\xi_\infty\in {\rm Ker}L.
\end{align*}
In other words, we have the following proposition.
\begin{prop}\label{sigular}
If ${\rm Ker}L=\{0\}$, any state in $\Sigma_\alpha(\omega)$
is $\omega$-singular i.e., $\omega$ is macroscopically unstable.
\end{prop}
%%%%%%%%%%%%%%%%%%
\subsection{\it The modular structure of the model}
Now we introduce the modular structure of our model.
As the structure is the same as that of \cite{JPNESS1},
we just state the results.
Let $\omega_S$ be a state over ${\cal O}_S$, defined by
$\omega_S={\rm Tr}(\rho_S\cdot)$ with a density matrix 
\begin{align*}
\rho_S=\sum_ip_i
\left\vert\psi_i\left\rangle\right\langle\psi_i\right\vert,
\end{align*}
on ${\mathfrak H}_S$.
Here $\{\psi_i\}$ are orthogonal unit vectors on ${\mathfrak H}_S$.
And let $\omega_\rho$ be a quasi-free state 
over ${\cal O}_f$
defined in (\ref{Fdis}).
Let $\bar A$ be a complex conjugation of $A$ with respect to
the orthogonal basis of ${\mathfrak H}_S$, that is
given by eigenvectors of $H_S$.
We denote the Fermi Fock space over ${\mathfrak h}\oplus{\mathfrak h}$
by ${\cal F}({\mathfrak h}\oplus{\mathfrak h})$.
Then we have the following proposition.
\begin{prop}
Suppose that $\omega_S$ is faithful, and $0<\rho<1$. 
The GNS triple $({\cal H}, \pi, \Omega)$ 
associated with the state
$\omega=\omega_S\otimes\omega_\rho$ is given by
\begin{align*}
{\cal H}={\cal H}_S\otimes{\cal H}_f,\quad
\Omega=\Omega_S\otimes\Omega_f,\quad
\pi=\pi_S\otimes\pi_f,
\end{align*}
where
\begin{align*}
&{\cal H}_S={\mathfrak H}_S\otimes{\mathfrak H}_S,\quad
\Omega_S=\sum p_i^{\frac12}\psi_i\otimes {\bar \psi_i},\quad
\pi_S\left(A\right)=A\otimes 1,\quad\quad A\in{\cal O}_S,\\
&{\cal H}_f={\cal F}\left({\mathfrak h}\oplus{\mathfrak h}\right),\quad
\Omega_f:{\rm vacuum\;vector\;of\;}{\cal F}
\left({\mathfrak h}\oplus{\mathfrak h}\right),\\
&\quad\pi_f\left(a\left(f\right)\right)
=a\left(\left(1-\rho\right)^{\frac 12}f\oplus 0\right)
+a^*\left(0\oplus \rho^{\frac 12}{\bar f}\right),\\
&\quad\pi_f\left(a^*\left(f\right)\right)
=a^*\left(\left(1-\rho\right)^{\frac 12}f\oplus 0\right)
+a\left(0\oplus \rho^{\frac 12}{\bar f}\right),\quad\quad f\in {\mathfrak h},
\end{align*}
and $\Omega$ is a cyclic and separating vector for the von Neumann
algebra $\pi({\cal O})^{''}$.
The dynamics $\alpha_t$ (\ref{cper}) defined on $C^*$-algebra
$\cal O$ extends to a weakly continuous one parameter
group of automorphisms $\tilde\alpha_t$
over $\pi({\cal O})^{''}$.
$\tilde\alpha_t$ is implemented by a strongly continuous one parameter
group of unitaries and the corresponding Liouville operator $L$
on $\cal H$ is
\begin{align}
&L=
\left(H_S\otimes 1\right)\otimes 1-
\left(1\otimes H_S\right)\otimes 1 +
1\otimes d\Gamma\left(\tilde h\right)\nonumber\\
&+\lambda\left(Y\otimes 1\right)
\otimes\left(
a\left(g_1\right)+a^*\left(g_1\right)\right)
-\lambda\left(1\otimes {\bar Y}\right)
\otimes\left(-1\right)^{N}
\left(a\left(g_2\right)-a^*\left(g_2\right)\right),
\label{defL}
\end{align}
where
\begin{align}
&\tilde h=h\oplus -h,
\label{defth}\\
&g_1=g_1^1\oplus g_1^2,\quad
g_2=g_2^1\oplus g_2^2,\\&\quad\;g_1^1=\left(1-\rho\right)^{\frac12}f,\quad
g_1^2=\rho^{\frac12}{\bar f},\nonumber\\
&\quad\;g_2^1=\rho^{\frac12}f,\quad
g_2^2=\left(1-\rho\right)^{\frac12}{\bar f}.
\label{gdef}
\end{align}
Here, $d\Gamma\left(\tilde h\right)$ is the second quantization of $\tilde h$.
\end{prop}
We denote the Fermion number operator on $\cal H$ by $N$.
We define the system Liouville operator $L_S$, the field Liouville operator $L_f$,
the free Liouville operator $L_0$, and
the interaction $I_0$ operator by
\begin{align}
&L_S=\left(H_S\otimes 1\right)\otimes 1-
\left(1\otimes H_S\right)\otimes 1,\nonumber\\
&L_f=1\otimes d\Gamma\left(\tilde h\right),\nonumber\\
&L_0=L_S+L_f,\nonumber\\
&I_0=
\left(Y\otimes 1\right)\otimes
\left(a\left(g_1\right)+a^*\left(g_1\right)\right)
-\left(1\otimes {\bar Y}\right)\otimes
\left(-1\right)^N\left(a\left(g_2\right)-a^*\left(g_2\right)\right).
\label{I0}
\end{align}
on $\cal H$.
%%%%%%%%%%%%%%%%%%%%%%%%%%%%%%%%%%%%%%%%%%%%%%%%%%%%%%%%%%%%%%%%%%%%%%%%%%
\section{Rescaling Group}\label{rescale}
In order to construct the positive commutator for the lattice system,
we introduce a strongly continuous one parameter group of
unitaries.
By the variable transformation $k\to t$, we have
\begin{align*}
{\mathfrak h}=L^2\left(\left[-\pi,\pi\right),{\rm d}k\right)
=L^2\left(\left(-\infty,\infty\right),{\rm d}\mu\right),
\end{align*}
where $\mu$ is the positive measure on $\mathbb R$ given by
\begin{align*}
{\rm d}\mu\left(t\right)=\frac{2}{t^2+1}{\rm d}t.
\end{align*}
The multiplication operator $m(k)$ on $\mathfrak{h}$ is transformed as
$m(k)\to m(k(t))$,
where $k(t)\equiv 2\tan^{-1}t$.
In particular, the single particle energy $h$ is transformed to
\begin{align*}
h\left(k\right)=\cos k-\gamma\quad
\to\quad
h\left(k\left(t\right)\right)
=\frac{2}{t^2+1}-1-\gamma.
\end{align*}
Now we introduce a unitary operator $u(\theta)$ 
on $\mathfrak h$ defined by
\begin{align*}
\left(u(\theta) f\right)\left(t\right)
\equiv{\rm e}^{\frac{\theta}{2}}
\sqrt{\frac{t^2+1}{{\rm e}^{2\theta}t^2+1}}
f\left({\rm e}^{\theta}t\right),
\quad \theta\in{\mathbb R},\quad
f\in L^2((-\infty,\infty),d\mu).
\end{align*}
By an elementary calculation, we have the following lemma;
\begin{lem}
$u(\theta)$ is a strongly continuous one parameter group of
unitaries.
\end{lem}
From this lemma, $u(\theta)$ is written as 
$u(\theta)={\rm e}^{i\theta p}$ with a selfadjoint operator $p$.
The action of $p$ on $C_0^{\infty}\left({\mathbb R}\right)$ is
\begin{align}
\left(pf\right)\left(t\right)
=-i\left(\frac 12 f\left(t\right)-\frac{t^2}{t^2+1}f\left(t\right)
+tf'\left(t\right)\right),
\quad f\in C_0^{\infty}\left({\mathbb R}\right).
\label{acp}
\end{align}
Let $\tilde u(\theta)$ be a unitary operator on 
${\mathfrak h}\oplus {\mathfrak h}$ defined by
\begin{align*}
\tilde u(\theta)\equiv u(\theta)\oplus u(-\theta).
\end{align*}
The generator of $\tilde u(\theta)$ is $\tilde p\equiv p\oplus -p$.
We define a unitary operator $U(\theta)$
on ${\cal H}$ 
by the second quantization $\Gamma(\tilde u(\theta))$
of $\tilde u(\theta)$:
\begin{align*}
U\left(\theta\right)\equiv 1\otimes
\Gamma\left(\tilde u\left(\theta\right)\right).
\end{align*}
Since $U(\theta)$ is a strongly continuous one parameter
unitary group, it is written as 
$U(\theta)={\rm e}^{\theta A_0}$,
where ${A_0}=1\otimes d\Gamma(i\tilde p)$ is an anti-selfadjoint operator.
%%%%%%%%%%%%%%%%%%%%%%%%%%%%%%%%%%%%%%%%%%%%%%%%%%%%%%%%

Now we introduce the commutator of $L_f$ and $A_0$
as a quadratic form on the dense set
$(D(N)\cap D(A_0))\;\times\;(D(N)\cap D(A_0))$.
By straight forward calculation,
the action of $\tilde u(\theta)$
on a multiplication operator $m_1\oplus m_2$
on ${\mathfrak h}\oplus {\mathfrak h}$ is
\begin{align*}
\left(\tilde u\left(-\theta\right)\left(m_1\oplus m_2\right)
\tilde u\left(\theta\right)\right)\left(t\right)
=m_1\left({\rm e}^{-\theta}t\right)\oplus
m_2\left({\rm e}^{\theta}t\right).
\end{align*}
Then, the action of $U(\theta)$ on
the second quantization $d\Gamma(m_1\oplus m_2)$
of $m_1\oplus m_2$ is
\begin{align}
U\left(-\theta\right)\left(
1\otimes{\rm d}\Gamma\left(m_1\oplus m_2\right)\right)
U\left(\theta\right)
&=1\otimes{\rm d}\Gamma\left(m_1^{-\theta}\oplus m_2^{\theta}\right),
\label{thetaaction}
\end{align}
where
\begin{align*}
(m_i^\theta f)\left(t\right)&=m_i\left({\rm e}^{\theta}t\right)f(t).
\end{align*}
%%%%%%%%%%%%%%%%%%%%%%%%%%%%%%%%%%%%%%%%%%%%%%%%%%%%%%%%%%%%%%%%%
We have the following proposition:
\begin{prop}\label{mulcom}
Suppose that $m_1,m_2$ are bounded
multiplication operators on $L^2({\mathbb R},d\mu)$
which are differentiable and satisfy
\begin{align*}
{\rm sup}_{t\in{\mathbb R}}\left\vert
qm_i(t)\right\vert<\infty,
\end{align*}
where 
\begin{align*}
qm_i\left(t\right)&\equiv-tm_i'\left(t\right).
\end{align*}
Then we have 
\begin{align}
\left\langle 1\otimes{\rm d}\Gamma\left(m_1\oplus m_2\right)\varphi,
A_0
\psi\right\rangle+
\left\langle A_0\varphi,
1\otimes{\rm d}\Gamma\left(m_1\oplus m_2\right)\psi\right\rangle
&=\left\langle \varphi,
1\otimes{\rm d}\Gamma\left(qm_1\oplus -qm_2\right)
\psi\right\rangle,
\label{commutatortheorem}
\end{align}
for all $\varphi,\psi\in D(N)\cap D(A_0)$.
\end{prop}
{\it Proof}\\
As $U(\theta)$ preserves number, we have $U(\theta)D(N)\subset D(N)$
and have the following relation for all $\varphi,\psi\in D(N)\cap D(A_0)$:
\begin{align}
&\lim_{\theta\to 0}\left\langle \varphi,
1\otimes\frac{{\rm d}\Gamma\left(m_1^{-\theta}\oplus m_2^\theta\right)-
{\rm d}\Gamma\left(m_1\oplus m_2\right)}
{\theta}
\psi\right\rangle\nonumber\\
&=\lim_{\theta\to 0}
\left\langle \left(1\otimes{\rm d}\Gamma\left(m_1\oplus m_2\right)\right)
U\left(\theta\right)\varphi,
\frac{U\left(\theta\right)-1}{\theta}\psi\right\rangle
+\left\langle \frac{U\left(\theta\right)-1}{\theta}\varphi,
1\otimes{\rm d}\Gamma\left(m_1\oplus m_2\right)\psi\right\rangle.
\label{comuproof}
\end{align}
Using dominated convergence theorem, we have 
the following statements:
\begin{enumerate}
\item For each $\psi\in D(N)$ and bounded operator $m$ on
$\mathfrak h\oplus\mathfrak h$, 
\begin{align*}
\lim_{\theta\to 0}\left(
1\otimes{\rm d}\Gamma\left(m\right)\right)
U\left(\theta\right)\psi
=\left(1\otimes{\rm d}\Gamma\left(m\right)\right)\psi,
\end{align*}
\item For each $\psi\in D(A_0)$,
\begin{align*}
\lim_{\theta\to 0}\frac{U\left(\theta\right)-1}{\theta}\psi
=A_0\psi,
\end{align*}
\item
For each $\psi\in D(N)$,
\begin{align*}
\lim_{\theta\to 0}
\left\Vert1\otimes\left(
\frac{{\rm d}\Gamma\left(m_1^{-\theta}\oplus m_2^\theta\right)-
{\rm d}\Gamma\left(m_1\oplus m_2\right)}
{\theta}-{\rm d}\Gamma\left(qm_1\oplus -qm_2\right)\right)
\psi\right\Vert
=0.
\end{align*}
\end{enumerate}
Substituting these to (\ref{comuproof}),
we obtain the statement of the proposition. $\square$
%%%%%%%%%%%%%%%%%%%%%%%%%%%%%%%%%%%%%%%%%%%%%%%%%%%

Substituting $m_1=h$, $m_2=-h$ to (\ref{commutatortheorem}),
we obtain 
\begin{align*}
\left\langle L_f\varphi,
A_0
\psi\right\rangle+
\left\langle A_0\varphi,L_f\psi\right\rangle
&=\left\langle \varphi,
S_1\psi\right\rangle,
\end{align*}
for all $\varphi,\psi\in D(N)\cap D(A_0)$.
Here $S_1$ is the second quantization 
$1\otimes {\rm d}\Gamma\left(s_1\oplus s_1\right)$ of $s_1\oplus s_1$ on
${\mathfrak h}\oplus{\mathfrak h}$, with $s_1$ defined by
\begin{align*}
\left(s_1f\right)(t)=s_1(t)f(t),\quad
s_1(t)=\frac{4t^2}{(1+t^2)^2}.
\end{align*}
Note that $S_1$ is positive.
This is the main point of this commutator.

Similarly, we have
\begin{align*}
\left\langle S_1\varphi,A_0\psi\right\rangle+
\left\langle A_0\varphi,S_1\psi\right\rangle
&=\left\langle \varphi,S_2\psi\right\rangle,\\
\left\langle S_2\varphi,
A_0
\psi\right\rangle+
\left\langle A_0\varphi,
S_2\psi\right\rangle
&=\left\langle \varphi,S_3\psi\right\rangle,
\end{align*}
where
\begin{align*}
&S_2=1\otimes{\rm d}\Gamma\left(s_2\oplus -s_2\right),\\
&\left(s_2f\right)(t)=s_2(t)f(t),\quad
s_2(t)=\frac{8\left(-1+t^2\right)t^2}{\left(1+t^2\right)^3},\\
&S_3=1\otimes{\rm d}\Gamma\left(s_3\oplus s_3\right),\\
&\left(s_3f\right)(t)=s_3(t)f(t),\quad
s_3(t)=\frac{16\left(t^2-4t^4+t^6\right)}{\left(1+t^2\right)^4}.
\end{align*}
Note that $S_1,S_2,S_3$ are all $N$-bounded.

The commutator of $A_0$ with the interaction term
can also be considered.
As the form factor $f\in{\mathfrak h}$ satisfies
$f\in{\cal D}(p^3)$,
$supp f\subset \Lambda_v$ and $\rho$ is differentiable in 
$\Lambda_v$, we have $g_1,g_2\in {\cal D}(\tilde p^3)$.
We get
\begin{align*}
\left\langle \varphi,I_0A_0\psi\right\rangle
+\left\langle A_0\varphi,I_0\psi\right\rangle
=\left\langle \varphi,I_1\psi\right\rangle,
\end{align*}
on $D(A_0)\times D(A_0)$.
Here $I_1$ is defined by
\begin{align*}
I_1&\equiv\left(Y\otimes 1\right)\otimes
\left(a\left(-i\tilde pg_1\right)+a^*\left(-i\tilde pg_1\right)\right)
-\left(1\otimes {\bar Y}\right)\otimes
\left(-1\right)^N\left(a\left(-i\tilde pg_2\right)
-a^*\left(-i\tilde pg_2\right)\right).
\end{align*}
Similarly, we have
\begin{align*}
\left\langle \varphi,I_1A_0\psi\right\rangle
+\left\langle A_0\varphi,I_1\psi\right\rangle
=\left\langle \varphi,I_2\psi\right\rangle,\quad
\left\langle \varphi,I_2A_0\psi\right\rangle
+\left\langle A_0\varphi,I_2\psi\right\rangle
=\left\langle \varphi,I_3\psi\right\rangle,
\end{align*}
with
\begin{align*}
I_2&\equiv\left(Y\otimes 1\right)\otimes
\left(a\left(-\tilde p^2g_1\right)+a^*\left(-\tilde p^2g_1\right)\right)
-\left(1\otimes {\bar Y}\right)\otimes
\left(-1\right)^N\left(a\left(-\tilde p^2g_2\right)-a^*\left(-\tilde p^2g_2\right)\right),\\
I_3&\equiv\left(Y\otimes 1\right)\otimes
\left(a\left(i\tilde p^3g_1\right)+a^*\left(i\tilde p^3g_1\right)\right)
-\left(1\otimes {\bar Y}\right)\otimes
\left(-1\right)^N\left(a\left(i\tilde p^3g_2\right)-a^*\left(i\tilde p^3g_2\right)\right).
\end{align*}
%%%%%%%%%%%%%%%%%%%%%%%%%%%%%%%%%%%%%%%%%%%%%%%%%%%
\section{Cut Off of the Interaction and Eigenvector}\label{cut}
In case the dispersion relation of the field $L_f$
is $\omega(k)=\vert k\vert$, 
$A_0$ is taken as the generator of the the shift operator.
And the commutator is given by the number operator $N$
: $[L_f,A_0]=N$.
Note that the dispersion relation of $N$ has a strictly positive
spectral gap $1>0$.
However in our case, the dispersion relation is
$\omega(k)=\cos(k)-\gamma$,
and the commutator is $[L_f,A_0]=S_1$.
Note that the dispersion relation of $S_1$ is 
positive, but it does not have a spectral gap:
it attains zero at $t=0$ and $t=\pm\infty$.
The existence of the spectral gap is essential in application of
positive commutator method as seen in the arguments in \cite{Merk01}.
So this situation causes a problem.
To overcome this difficulty, we assume the Assumption \ref{suppg}
on the interaction.
In this section, we see that as the result of the cut off
$supp f\subset \Lambda_v$,
any eigenvector of $L$ is in the range of $P=P(N_{\Lambda_v^c}=0)$.

\newtheorem{kthm}{Assumption\ref{suppg}}
\renewcommand{\thekthm}{}
Now let $\tilde{\mathfrak h}={\mathfrak h}\oplus{\mathfrak h}$.
We decompose $\tilde{\mathfrak h}$
with respect to ${\mathbb R}=\Lambda_v\oplus \Lambda_v^c$:
\begin{align*}
\tilde{\mathfrak h}&=
{\tilde {\mathfrak h}}_{\Lambda_v}\oplus
{\tilde {\mathfrak h}}_{\Lambda_v^c},\\
&{\tilde {\mathfrak h}}_{\Lambda_v}
=L^2(\Lambda_v,d\mu)\oplus L^2(\Lambda_v,d\mu),\quad
{\tilde {\mathfrak h}}_{\Lambda_v^c}
=L^2(\Lambda_v^c,d\mu)\oplus L^2(\Lambda_v^c,d\mu).
\end{align*}
Note that $\tilde{h}$ (\ref{defth}) is decomposed into
$\tilde{h}=
{\tilde {h}}_{\Lambda_v}\oplus
{\tilde {h}}_{\Lambda_v^c},$
with respect to this decomposition of the Hilbert space,
because it is a multiplication operator.
Let $N_{\Lambda_v}$, $N_{\Lambda_v^c}$ be
\begin{align*}
N_{\Lambda_v}\equiv 1\otimes {\rm d}\Gamma(1_{\Lambda_v}\oplus 0),
\quad
N_{\Lambda_v^c}\equiv 1\otimes{\rm d}\Gamma(0\oplus 1_{\Lambda_v^c}),
\end{align*}
on ${\cal H}$ with respect to this decomposition.
Below for a Hilbert space ${\mathfrak K}$,
we denote by ${\cal U}({\mathfrak K})$
a set of unitary operators on ${\mathfrak K}$,
and by ${\cal F}({\mathfrak K})$ the Fock space over $\mathfrak K$.
Furthermore, $\Gamma(u)$ is the second quantization of
$u\in{\cal U}({\mathfrak K})$, and for a self-adjoint
operator ${\kappa}$ on $\mathfrak K$,
$d\Gamma_{\mathfrak K}({\kappa})$ is the second quantization of 
$\kappa$.
We have the following proposition:
\begin{prop}
There is a unitary operator 
$U:{\cal F}({\tilde {\mathfrak h}}_{\Lambda_v}\oplus
{\tilde {\mathfrak h}}_{\Lambda_v^c})
\to {\cal F}({\tilde {\mathfrak h}}_{\Lambda_v})
\otimes {\cal F}({\tilde {\mathfrak h}}_{\Lambda_v^c})$
which satisfies the following conditions;
\begin{enumerate}
\item For any $u_{\Lambda_v}
\in {\cal U}({\tilde {\mathfrak h}}_{\Lambda_v})$
and
$u_{\Lambda_v^c}\in {\cal U}({\tilde {\mathfrak h}}_{\Lambda_v^c})$,
\[
U\Gamma(u_{\Lambda_v}\oplus u_{\Lambda_v^c})U^*
=\Gamma(u_{\Lambda_v})
\otimes\Gamma(u_{\Lambda_v^c}),\]
\item For any $f\in {\tilde {\mathfrak h}}_{\Lambda_v}$ and
$g\in{\tilde {\mathfrak h}}_{\Lambda_v^c}$
\[ Ua(f\oplus g)U^*=a(f)\otimes 1+(-1)^{N_{\Lambda_v}}\otimes a(g).\]
\end{enumerate}
\label{compunitary}
\end{prop}
By this proposition, we have
\begin{align*}
U\Gamma({\rm e}^{it\tilde h_{\Lambda_v}}\oplus
{\rm e}^{it\tilde h_{\Lambda_v^c}})U^*
=\Gamma({\rm e}^{it\tilde h_{\Lambda_v}})
\otimes\Gamma({\rm e}^{it\tilde h_{\Lambda_v^c}}).
\end{align*}
Recall that $L$ is given by (\ref{defL}).
As the Assumption \ref{suppg} $supp f\subset \Lambda_v$ implies 
$g_1,g_2\in\tilde{\mathfrak h}_{\Lambda_v}$,
we obtain the following unitary equivalence;
\begin{multline*}
ULU^*
=L_S
+1\otimes {\rm d}\Gamma_{\Lambda_v}(\tilde h_{\Lambda_v})\otimes 1
+1\otimes 1\otimes {\rm d}\Gamma_{\Lambda_v^c}(\tilde h_{\Lambda_v^c})\\
+\lambda\left(Y\otimes 1\right)\otimes
\left(a\left(g_1\right)+a^*\left(g_1\right)\right)\otimes 1
-\lambda\left(1\otimes {\bar Y}\right)\otimes
\left(-1\right)^{N_{\Lambda_v}}
\left(a\left(g_2\right)-a^*\left(g_2\right)\right)\otimes 
\left(-1\right)^{N_{\Lambda_v^c}}.
\end{multline*}
By this equivalence, we have the following lemma,
\begin{lem}
$L$ and $N_{\Lambda_v^c}$ strongly commute.
\label{strongcom}
\end{lem}
Similarly, $N_{\Lambda_v^c}$ strongly commutes
with $L_S$, $L_f$, $L_0$, $\lambda I_0$
$N$, $S_1$.
Using this fact, we obtain the following proposition:
\begin{prop}
Suppose $\psi$ is an eigenvector of $L$.
Then $\psi$ satisfies
\[P\left(N_{\Lambda_v^c}=0\right)\psi=\psi,\]
where $P\left(N_{\Lambda_v^c}=0\right)$ is the spectral 
projection of $N_{\Lambda_v^c}$
corresponding to $N_{\Lambda_v^c}=0$.
\end{prop}
{\it Proof}\\
Let $P_e$ (resp.$P_o$) be the spectral projection 
of $N_{\Lambda_v^c}$ onto
$N_{\Lambda_v^c}={\rm even}$ (resp.$N_{\Lambda_v^c}={\rm odd}$).
Let us decompose the Hilbert space $\cal H$ as
\begin{align*}
{\cal H}=P_e{\cal H}\oplus P_o{\cal H}.
\end{align*}
Then by Lemma \ref{strongcom}, $L$ is decomposed into
\begin{align}
L=P_eL\oplus P_oL,
\label{Ldecomposition}
\end{align}
with respect to
the subspaces $P_e{\cal H}$ and $ P_o{\cal H}$.
In particular, if $\psi$ is an eigenvector of $L$,
$P_e\psi$ is an eigenvector of $P_eL$ and
$P_o\psi$ is an eigenvector of $P_oL$.
With respect to the decomposition
${\cal H}={\cal H}_S\otimes{\cal F}({\tilde {\mathfrak h}}_{\Lambda_v})
\otimes {\cal F}({\tilde {\mathfrak h}}_{\Lambda_v^c})$,
$P_eL$ on $P_e{\cal H}$ is 
\begin{align*}
\left. P_eL\right\vert_{P_e{\cal H}}=\left(
L_S +
1\otimes{\rm d}\Gamma_{\Lambda_v}\left(\tilde h_{\Lambda_v}\right)
+V_1+V_2\right)\otimes 1
+(1\otimes 1)\otimes {\rm d}\Gamma_{\Lambda_v^c}
\left(\tilde h_{\Lambda_v^c}\right)\vert_{P_e{\cal H}}
\end{align*}
with
\begin{align*}
V_1&=\lambda\left(Y\otimes 1\right)\otimes
\left(a\left(g_1\right)+a^*\left(g_1\right)\right)\\
V_2&=-\lambda\left(1\otimes {\bar Y}\right)\otimes
\left(-1\right)^{N_{\Lambda_v}}
\left(a\left(g_2\right)-a^*\left(g_2\right)\right),
\end{align*}
because
\begin{align*}
P_e=1\otimes P(N_{\Lambda_v^c}=even).
\end{align*}
On the other hand,
$P_oL$ on $P_o{\cal H}$ is 
\begin{align*}
\left.P_oL\right\vert_{{\cal H}_o}=\left(
L_S+
1\otimes {\rm d}\Gamma_{\Lambda_v}\left(\tilde h_{\Lambda_v}\right)
+V_1-V_2\right)\otimes 1
+(1\otimes 1)\otimes {\rm d}\Gamma_{\Lambda_v^c}(\tilde h_{\Lambda_v^c})
\vert_{P_o{\cal H}}
\end{align*}
because
\begin{align*}
P_o=1\otimes P(N_{\Lambda_v^c}=odd).
\end{align*}
Note that ${\rm d}\Gamma_{\Lambda_v^c}(\tilde h_{\Lambda_v^c})$ and 
$N_{\Lambda_v^c}$ strongly commute
on ${\cal F}(\tilde h_{\Lambda_v^c})$.So
${\rm d}\Gamma_{\Lambda_v^c}(\tilde h_{\Lambda_v^c})$
is decomposed into
$P_e{\rm d}\Gamma_{\Lambda_v^c}(\tilde h_{\Lambda_v^c})
\oplus P_o{\rm d}\Gamma_{\Lambda_v^c}(\tilde h_{\Lambda_v^c})$.
On the other hand,
${\rm d}\Gamma_{\Lambda_v^c}(\tilde h_{\Lambda_v^c})$ in 
${\cal F}({\tilde {\mathfrak h}}_{\Lambda_v^c})$ has 
unique eigenvector $\Omega_{\Lambda_v^c}$
which is the vacuum vector of ${\cal F}({\tilde {\mathfrak h}}_{\Lambda_v^c})$.
So $\left.{\rm d}\Gamma_{\Lambda{_v^c}}
\right\vert_{P_e{\cal F}({\tilde {\mathfrak h}}_{\Lambda_v^c})}$
has the unique eigenvector $\Omega_{\Lambda_v^c}$,
while $\left.{\rm d}\Gamma_{\Lambda_v^c}
\right\vert_{P_o{\cal F}({\tilde {\mathfrak h}}_{\Lambda_v^c})}$
has no eigenvector.
Hence by Theorem \ref{eigentensor}, if there exists
an eigenvector $\psi_e$
of $P_eL$, it is of the form
\begin{align*}
\psi_e=\varphi\otimes\Omega_{\Lambda_v^c},
\end{align*}
and $P_oL$ has no eigenvector.
Hence if $\psi$ is the eigenvector of $L$,
we have
\begin{align*}
P_e\psi=\varphi\otimes\Omega_{\Lambda_v^c},\;\;\;
P_o\psi=0.
\end{align*}
That is,
\begin{align*}
\psi=\varphi\otimes\Omega_{\Lambda_v^c}.
\end{align*}
This means
\begin{align*}
P(N_{\Lambda_v^c}=0)\psi=\varphi\otimes 
P(N_{\Lambda_v^c}=0)\Omega_{\Lambda_v^c}
=\varphi\otimes\Omega_{\Lambda_v^c}=\psi.
\quad\quad\quad\quad\quad\quad\square
\end{align*}
%%%%%%%%%%%%%%%%%%%%%%%%%%%%%%%%%%%%%%%%%%%%%%%%%%%%%%%%%%%%%%%%%%%%%%%
\section{Positive Commutator}\label{PC}
In this section, we construct strictly positive commutator.
First we define the operator $\left[L,A_0\right]$ on
$D(\left[L,A_0\right])\equiv D(N)$ by
\begin{align*}
\left[L,A_0\right]\equiv S_1+\lambda I_1.
\end{align*}
Note that this is defined as an operator of the right hand side,
and not as a commutator $LA_0-A_0L$.
However the arguments in Section \ref{rescale} guarantees
\begin{align*}
\left\langle L\varphi,A_0\psi\right\rangle+
\left\langle A_0\varphi,L\psi\right\rangle
=\left\langle \varphi,(S_1+\lambda I_1)\psi\right\rangle
=\left\langle \varphi,\left[L,A_0\right]\psi\right\rangle,
\end{align*}
for all $\psi,\varphi\in D(A_0)\cap D(N)$.

As seen in the previous section, we consider the subspace $P{\cal H}$.
If we restrict ourselves to $P(1\otimes P_{\Omega_f})^{\perp}$,
we have
\begin{align*}
P(1\otimes P_{\Omega_f})^{\perp}[L,A_0](1\otimes P_{\Omega_f})^{\perp}P
\ge (v-\lambda\Vert I_1\Vert)P(1\otimes P_{\Omega_f})^{\perp},
\end{align*}
and $[L,A_0]$ is strictly positive on 
$P(1\otimes P_{\Omega_f})^{\perp}{\cal H}$
for $\lambda$ small enough.
But for $P(1\otimes P_{\Omega_f}){\cal H}$, we have
\begin{align*}
P(1\otimes P_{\Omega_f})[L,A_0](1\otimes P_{\Omega_f})P=0.
\end{align*}
So we need to modify the commutator to make it strictly positive
on $P(1\otimes P_{\Omega_f}){\cal H}$, too.
This is done by introducing a bounded operator $b$:
\begin{align*}
b=b(e)=\theta\lambda(\bar QR_{\epsilon}^2I_0Q
-QI_0R_{\epsilon}^2\bar Q),\\
R_{\epsilon}=R_{\epsilon}(e)=\left(\left(L_0-e\right)^2+\epsilon^2\right)^{-\frac12},
\end{align*}
where $Q\equiv P(L_S=e)\otimes P_{\Omega_f},\quad
{\bar Q}\equiv 1-Q$ and $I_0$ is defined at (\ref{I0}).
The parameters $\theta$ and $\epsilon$ will be determined later.
We denote $R_{\epsilon}^2\bar Q$
by $\bar{R_{\epsilon}^2}$.
We define $[L,b(e)]$ on $D(N)$ by $[L,b(e)]\equiv Lb(e)-b(e)L$.
Note that $L_Sb(e)-b(e)L_S$ and $\lambda I_0b(e)-b(e)\lambda I_0$ are bounded.
For all $\psi\in D(N)$,
\begin{align*}
[L_f,b(e)]\psi
=\theta\lambda\left(\bar QR_{\epsilon}^2
\left[L_f,I_0\right]Q-Q\left[L_f,I_0\right]R_{\epsilon}^2\bar Q
\right)\psi
=\theta\lambda\left(\bar QR_{\epsilon}^2
\tilde I_1 Q-Q\tilde I_1R_{\epsilon}^2\bar Q
\right)\psi,
\end{align*}
where 
\begin{align*}
\tilde I_1\equiv
\left(Y\otimes 1\right)\otimes
\left(-a\left(\tilde h g_1\right)+a^*\left(\tilde hg_1\right)\right)
+\left(1\otimes {\bar Y}\right)\otimes
\left(-1\right)^N\left(a\left(\tilde hg_2\right)
+a^*\left(\tilde hg_2\right)\right).
\end{align*}
As $\tilde I_1$ is bounded, it can be extended to the whole $\cal H$.
We denote 
the extension by the same symbol $[L,b(e)]$.
We define $[L,A]$ on $D(N)$
by
\begin{align*}
\left[L,A\right]=\left[L,A_0\right]+\left[L,b\right]
=S_1+\lambda I_1+\left[L,b\right].
\end{align*}
By the above choice of $b$, we have now
\begin{align*}
&P(1\otimes P_{\Omega_f})[L,A](1\otimes P_{\Omega_f})P\\
&=2\theta\lambda^2P(P(L_S=e)\otimes P_{\Omega_f})
I_0\bar R_{\epsilon}^2I_0
(P(L_S=e)\otimes P_{\Omega_f})P\neq 0.
\end{align*}
In this section, we prove the following theorem:
\begin{thm}\label{positivecommutator}
Let $\Delta$ be the interval of ${\mathbb R}$
whose interior includes an eigenvalue $e$ of $L_S$,
and no other eigenvalue is included in $\bar\Delta$, i.e, 
$\sigma(L_S)\cap\bar\Delta=\{e\}$.
Let $\varsigma\in C_0^\infty({\mathbb R})$
be a smooth function s.t.$\varsigma=1$ on $\Delta$
and ${\rm supp}\; \varsigma\cap\sigma(L_S)=\{e\}$.
Let $\theta$ and $\epsilon$ be
\[
\epsilon=\lambda^{\frac{44}{100}},\quad
\theta=\lambda^{\frac{26}{100}}.
\]
Let $P$ be the spectral projection of $N_{\Lambda_v^c}$
onto $\{0\}$.
Suppose that Assumption \ref{suppg} and \ref{O} are satisfied.
Then there exists $\lambda_1>0$ s.t.
\begin{align}
P\varsigma(L)\left[L,A\right]\varsigma(L)P
\ge
\frac{1}{2}\cdot\lambda^{\frac{182}{100}}\cdot\gamma_e
P\varsigma(L)\left(1-14\tilde P_e\right)\varsigma(L)P,
\label{final}
\end{align}
for any $0< \lambda< \lambda_1$.
Furthermore,
if we assume $\beta_0<\beta_+,\beta_-<\beta_1$,
$\Vert p f\Vert,\Vert f\Vert\le b$ for any fixed 
$0<\beta_0<\beta_1<\infty$ and $0<b<\infty$,
we can choose $\lambda_1$ as
\[
\lambda_1=C\min\{
v^{\frac{100}{26}},\left(\frac{v}{\gamma_e}\right)^{\frac{100}{182}},
(\gamma_e)^{\frac{100}{11}},(v\gamma_e)^{\frac{100}{18}}
\}.
\]
Here $C$ is a constant which depends on $\beta_0,\beta_1$ and $b$, but
is independent of $v$ and $\beta_+-\beta_-$.
\end{thm}
We carry out the proof in two steps.
First, we show the strict positivity of $[L,A]$
with respect to the spectral localization in $L_0$.
Second  we derive the strict positivity of $[L,A]$ 
with respect to the spectral localization in $L$.
%%%%%%%%%%%%%%%%%%%%%%%%%%%%%%%%%%%%%%%%%%%%%%%%%
\subsection{\it Positive commutator with respect to the 
spectral localization in $L_0$}
In this subsection, we take the first step.
Here is the main theorem of this subsection;
\begin{thm}\label{fc}
Let $\Delta$ be an interval of ${\mathbb R}$,
containing exactly one eigenvalue $e$ of $L_S$
and let $E_\Delta^0$ be the spectral projection of 
$L_0$ onto $\Delta$.
Suppose that Assumption \ref{suppg} and \ref{O} are satisfied.
Then there exists $0<t_i,\;i=1,\cdots 5$ such that,
if $\lambda,\;\epsilon,\;\theta$ satisfy
\[
\lambda+\theta\lambda^2\epsilon^{-2}<t_1,\quad
\theta<t_2,\quad
\theta\lambda^2\epsilon^{-1}<t_3,\quad
\epsilon<t_4,\quad
\theta^{-1}\epsilon+\theta\lambda^2\epsilon^{-3}<t_5,
\]
then 
\[
PE_\Delta^0[L,A]E_\Delta^0P\ge
\frac{\theta\lambda^2}{\epsilon}
\gamma_e\left(1-7{\tilde P}_e
\right)E_\Delta^0P.\]
\label{L0bound}
\end{thm}
The proof goes parallel to \cite{Merk01}.
The difference emerges from the difference of the commutator $[L_0,A_0]$:
in \cite{Merk01}, it was $[L_0,A_0]=N$, while we have $[L_0,A_0]=S_1$.
Whereas $N$ has a spectral gap, $S_1$ has no spectral gap.
Since positive commutator method makes use of the finite spectral gap,
the lack of the gap causes a difficulty.
To overcome this, we introduced the cut off in Section \ref{cut}.

To prove the theorem, we use the Feshbach map theorem \cite{BFS98};
\begin{thm}
Let $H$ be a closed operator densely defined on 
a Hilbert space $\cal H$,
and let $P$ be a projection operator such that ${\rm Ran}P\subset D(H)$.
We set $\bar{P}=1-P$.
Then $H_{\bar P}\equiv {\bar P}H{\bar P}$ is a densely defined 
operator on ${\bar P}{\cal H}$.
Let $z$ be an element in
the resolvent set $\rho(H_{\bar P})$ of $H_{\bar P}$
on ${\bar P}{\cal H}$. Assume
\[
\left\Vert PH{\bar P}\left(H_{\bar P}-z\right)^{-1}{\bar P}\right\Vert
<\infty,\quad
\left\Vert 
{\bar P}\left(H_{\bar P}-z\right)^{-1}{\bar P}HP
\right\Vert
<\infty.
\]
Then the Feshbach map
\[
f_P(H-z)\equiv P(H-z)P-PH{\bar P}(H_{\bar P}-z)^{-1}{\bar P}HP
\]
is well defined on $P\cal H$.
$f_P(H-z)$ and $H$ have the isospectral property in the sense that
\[
z\in\sigma(H)\iff 0\in\sigma(f_P(H-z)),\quad
z\in\sigma_{PP}(H)\iff 0\in\sigma_{PP}(f_P(H-z)).
\]
Here $\sigma$ and $\sigma_{PP}$ represent spectrum and
pure point spectrum respectively.
\label{feshbach}
\end{thm}
To apply the theorem, we introduce 
\begin{align*}
\chi_\nu\equiv P(N\le\nu).
\end{align*}
By this $\chi_\nu$, we can consider $B_{ij}'$ below
as bounded operators.
Projections $\tilde{P}_e,\chi_\nu,P,Q,E_\Delta^0$
are in relation that
\begin{align*}
{\tilde P}_{e}\le\chi_\nu,\quad
{\tilde P}_{e}\le P,\quad
{\tilde P}_{e}\le Q,\quad
Q\le E_\Delta^0,\quad Q\le P.
\end{align*}
Hence we have the followings; 
\begin{align*}
{\bar Q}{\tilde P}_{e}=0,\quad
{\bar Q}{\tilde P}_{e}^{\perp}={\bar Q},\quad
E_\Delta^0P\chi_\nu{\tilde P}_{e}={\tilde P}_{e}.
\end{align*}
As $L_0$, $N_{\Lambda_v^c}$, $N$, $L_S$ strongly commute,
the projections $E_\Delta^0$, $Q$, $P$, $\chi_\nu$ commute each other.
So we can define the following projections
\begin{align}
Q_1\equiv QE_\Delta^0P\chi_\nu,\quad Q_2\equiv {\bar Q}E_\Delta^0P\chi_\nu,
\label{Qdef}
\end{align}
which satisfy
\begin{align}
Q_1{\tilde P}_e={\tilde P}_e,\quad
Q_2{\tilde P}_e=0,\quad
Q_2{\tilde P}_e^{\perp }=Q_2.
\label{Q1Q2}
\end{align}
The relation
$NP=N_{\Lambda_v}P$ implies 
\begin{align*}
N_{\Lambda_v}Q_2=N_{\Lambda_v}P{\bar Q}E_\Delta^0\chi_\nu
=NP{\bar Q}E_\Delta^0\chi_\nu=NQ_2.
\end{align*}
As $P(N=0)=1\otimes P_{\Omega_f}$,
and $\Delta\cap\sigma(L_0)=\{e\}$, we have
\begin{align*}
P(N=0)\bar Q
=\left(1-P(L_S=e)\right)\otimes P_{\Omega_f}
=\sum_{f\neq e}P(L_S=f)\otimes P_{\Omega_f}
\le P(L_0\subset \Delta^c),
\end{align*}
which implies
\begin{align*}
P(N=0)Q_2=P(N=0)\bar QE_\Delta^0P\chi_\nu=0.
\end{align*}
It follows that
\begin{align}
NQ_2=N_{\Lambda_v}Q_2\ge Q_2.
\label{NonQ2}
\end{align}
{\it Proof of Theorem \ref{L0bound}}\\
We apply Theorem {\ref {feshbach}} to the operator
$B'$ defined by
\begin{align*}
B'\equiv [L,A]-\delta_e\cdot{\tilde P}_e^{\perp }.
\end{align*}
Here, $\delta_e$ is a parameter which will be determined later.
We define bounded operators $B_{ij}'$ by
\begin{align*}
B_{ij}'\equiv Q_iB'Q_j\quad i,j=1,2.
\end{align*}
If $\vartheta\in \rho(B_{22}')$, then
\begin{align*}
\varepsilon\left({\vartheta}\right)
\equiv B_{11}'-B_{12}'(B_{22}'-{\vartheta})^{-1}B_{21}',
\end{align*}
is well-defined and 
\begin{align*}
\varepsilon\left({\vartheta}\right)
=f_{Q_1}\left(B'-\vartheta\right)+\vartheta Q_1.
\end{align*}
First we investigate the lower bound of $B_{22}'$ in order to
study the resolvent set $\rho(B_{22}')$.
\begin{prop}\label{pB22}
We have
\[
B_{22}'\ge \left(v
-\delta_e-C_1\left(\lambda+\theta\lambda^2\epsilon^{-2}\right)
\right)
\cdot Q_2,
\]
where 
\begin{align*}
C_1=\Vert I_1\Vert
+2\cdot\Vert I_0\Vert^2.
\end{align*}
\end{prop}
{\it Proof}\\
We estimate each term of
\begin{align*}
B_{22}'=Q_2(S_1+\lambda I_1+\left[L,b\right]
-\delta_e\cdot \tilde P_e^{\perp}
)Q_2.
\end{align*}
From the inequality $S_1\ge v\cdot N_{\Lambda_v}$ and
the inequality (\ref{NonQ2}), the first term is bounded below as
$Q_2S_1Q_2\ge vQ_2$.
The lower bound of
the second term is given by
$\lambda I_1
\ge -\lambda\cdot \Vert I_1\Vert$, as $I_1$ is bounded.
Let us estimate the norm of the third term.
Substituting $L=L_0+\lambda I_0$, $L_0Q=QL_0$
and $Q_2Q=0$, we obtain
$Q_2\left[L,b \right]Q_2
=-\theta\lambda^2Q_2(I_0QI_0R_\epsilon^2\bar Q+\bar QR_\epsilon^2 I_0QI_0)Q_2.$
Then it is estimated as
$\Vert Q_2\left[L,b \right]Q_2\Vert
\le 2\cdot \Vert I_0\Vert^2{\theta\lambda^2}{\epsilon^{-2}}$,
since we have 
$\Vert R_\epsilon^2\Vert\le\epsilon^{-2}$.
The fourth term is $Q_2{\tilde P}_e^{\perp}Q_2=Q_2$
by (\ref{Q1Q2}).
Hence we obtain
\begin{align*}
B_{22}'\ge
\left(v-C_1\left(\lambda+\theta\lambda^2\epsilon^{-2}\right)
-\delta_e
\right)Q_2
\end{align*}
where
$C_1=\Vert I_1\Vert+2\cdot \Vert I_0\Vert^2.
\quad\quad\square$\\
Suppose $\lambda,\theta,\epsilon$ satisfies 
\begin{align*}
\lambda+\theta\lambda^2\epsilon^{-2}<t_1,
\end{align*}
where
$t_1=\frac{v}{5C_1}.$
Then if $\vartheta$ satisfies 
$\vartheta\le\frac 12v-\delta_e$,
it is in the resolvent set $\rho(B_{22}')$ of $B_{22}'$ by this proposition.

Next we estimate the lower bound of $\varepsilon\left({\vartheta}\right)$.
\begin{prop}\label{epsi}
Suppose $\lambda,\theta,\epsilon$ satisfies 
\begin{align*}
\lambda+\theta\lambda^2\epsilon^{-2}<t_1,
\end{align*}
where
$t_1=\frac{v}{5C_1}.$
Then if $\vartheta$ satisfies 
$\vartheta\le\frac 12v-\delta_e$,
it is in the resolvent set $\rho(B_{22}')$ of $B_{22}'$ and
the following inequality holds;
\begin{multline*}
\varepsilon\left({\vartheta}\right)
\ge 2\theta\lambda^2Q_1\left[I_0\bar R_{\epsilon}^2I_0
-\frac{1}{2\theta\lambda^2}\delta_e\cdot {\tilde P}_e^{\perp}
\right]Q_1
-C_2\left(\theta^2\lambda^2 Q_1I_0 \bar R_\epsilon^2 I_0Q_1
+(\lambda^2+\theta^2\lambda^4\epsilon^{-4})Q_1
\right)
\end{multline*}
where
$C_2=10v^{-1}\Vert I_1\Vert^2
+20v^{-1}+80v^{-1}\Vert I_0\Vert^4.$
\label{epsilon}
\end{prop}
{\it Proof}\\
First $B_{11}'$ is given by
\begin{align*}
B_{11}'=Q_1\left(S_1+\lambda I_1+\left[L,b\right]
-\delta_e\cdot \tilde P_e^{\perp}\right)Q_1.
\end{align*}
As $S_1Q=0$ and $aQ=Qa^*=0$, the first and the second term vanish.
Using $\bar QQ_1=0$, $L_0Q=QL_0$, we obtain
%\begin{align*}
%Q_1[L,b]Q_1=2\theta\lambda^2Q_1I_0\bar R_{\varepsilon}^2I_0Q_1
%\quad \bar R_{\varepsilon}^2=\bar QR_{\varepsilon}^2.
%\end{align*}
%Hence we obtain
\begin{align}
B_{11}'=2\theta\lambda^2Q_1\left[I_0\bar R_{\epsilon}^2I_0
-\frac{1}{2\theta\lambda^2}\delta_e\cdot {\tilde P}_e^{\perp}
\right]Q_1.
\label{B11}
\end{align}
Second, 
$Q_2\left(B_{22}'-\vartheta\right)^{-1}Q_2$
is evaluated as follows;
\begin{lem}
Suppose
$\lambda+\theta\lambda^2\epsilon^{-2}<t_1$.
Then for all $\vartheta\le\frac 12v-\delta_e$,
$\vartheta$ is in the resolvent set $\rho(B_{22}')$ of $B_{22}'$
and
\[
0\le \left\langle\psi, 
Q_2\left(B_{22}'-\vartheta\right)^{-1}Q_2
\psi\right\rangle
\le 5\Vert S_1^{-\frac12}Q_2\psi\Vert^2,
\quad
\forall \psi\in\cal H.\]
\label{B22}
\end{lem}
{\it Proof}\\
As $S_1Q_2\ge vN_{\Lambda_v}Q_2\ge vQ_2$,
$S_1$ is invertible on $Q_2{\cal H}$.
The proof is similar to that of equation (41) in \cite{Merk01}.
In facts, it is easier because we are considering Fermion system,
whose interaction terms are bounded.
We omit the details.$\square$\\
Third, $\left\Vert S_1^{-\frac12}B_{21}'\psi\right\Vert$
is estimated as follows;
\begin{lem}
\[
\left\Vert S_1^{-\frac12}B_{21}'\psi\right\Vert^2
\le C_2'
\left(
\theta^2\lambda^2\Vert\bar R_{\epsilon} I_0Q_1\psi\Vert^2
+\theta^2\lambda^4\epsilon^{-4}\Vert \psi\Vert^2
+\lambda^2\Vert \psi\Vert^2
\right)
\]where
$C_2'=
2v^{-1}\Vert I_1\Vert^2
+4v^{-1}+16v^{-1}\Vert I_0\Vert^4.$
\label{B21}
\end{lem}
{\it Proof}\\
The proof is the same as that of \cite{Merk01} (p342).$\square$

Now let us complete the proof of Proposition \ref{epsilon}.
Combining Lemma \ref{B22} and \ref{B21},
we obtain the following evaluation:
if $\lambda,\theta,\epsilon$ satisfies
$\lambda+\theta\lambda^2\epsilon^{-2}<t_1$,
then 
for all $\vartheta\le\frac 12v-\delta_e$, we have
$\vartheta\in\rho(B_{22}')$ and
\begin{align*}
0\le \left\langle\psi,
B_{12}'\left(B_{22}'-\vartheta\right)^{-1}B_{21}'
\psi\right\rangle
\le
C_2\left(\theta^2\lambda^2 \left\Vert\bar R_\epsilon I_0Q_1\psi\right\Vert^2
+(\lambda^2+\theta^2\lambda^4\epsilon^{-4})\Vert\psi\Vert^2
\right)
\end{align*}
where
$C_2=10v^{-1}\Vert I_1\Vert^2
+20v^{-1}+80v^{-1}\Vert I_0\Vert^4$.
Combining this with (\ref{B11}), we obtain the required estimation.
$\square$

Now let us complete the proof of Theorem \ref{fc}.
Suppose that  $\lambda+\theta\lambda^2\epsilon^{-2}<t_1$.
Then by Proposition \ref{epsi},
for all $\vartheta\le\frac 12v-\delta_e$,
we have $\vartheta\in\rho(B_{22}')$ and
\begin{align*}
\varepsilon\left({\vartheta}\right)
\ge D,
\end{align*}
where
\begin{align*}
D=2\theta\lambda^2Q_1\left[I_0\bar R_{\epsilon}^2I_0
-\frac{1}{2\theta\lambda^2}\delta_e\cdot {\tilde P}_e^{\perp}
\right]Q_1
-C_2\left(\theta^2\lambda^2 Q_1I_0 \bar R_\epsilon^2 I_0Q_1
+(\lambda^2+\theta^2\lambda^4\epsilon^{-4})Q_1
\right).
\end{align*}
Note that $D$ is $\vartheta$-independent.

Now let $\vartheta_0\equiv \inf\sigma( B'\vert_{(Q_1\oplus Q_2)\cal H})$.
Here we have two cases,
\begin{enumerate}
\item $\vartheta_0\ge\frac 12v-\delta_e$.
\item $\frac 12v-\delta_e>\vartheta_0$.
In this case, $\vartheta_0\in\rho(B_{22}')$ and
$\varepsilon\left({\vartheta_0}\right)$ is well-defined.
Then $f_{Q_1}(B'-\vartheta_0)$ satisfies
\begin{align*}
f_{Q_1}(B'-\vartheta_0)
=\varepsilon\left({\vartheta_0}\right)-\vartheta_0\cdot Q_1
\ge \left(\inf{\sigma(D)}-\vartheta_0\right)\cdot Q_1.
\end{align*}
On the other hand, note that
$\vartheta_0\in \sigma(B'\vert_{(Q_1\oplus Q_2)\cal H})$.
By Feshbach Theorem \ref{feshbach}, this implies
$0\in\sigma(f_{Q_1}(B'-\vartheta_0))$.
Hence $f_{Q_1}(B'-\vartheta_0)$ is not invertible, and we get
\begin{align*}
\vartheta_0\ge\inf{\sigma(D)}.
\end{align*}
\end{enumerate}
So we have
\begin{align*}
\vartheta_0\ge\min\{\frac 12v-\delta_e,\inf{\sigma(D)}\},
\end{align*}
i.e.,
\begin{align}
B'\vert_{(Q_1\oplus Q_2)\cal H}\ge
\min\{\frac 12v-\delta_e,\inf{\sigma(D)}\}.
\label{Bd}
\end{align}

The next problem is to investigate the lower bound of $D$.
%%%%%%%%%%%%%%%%%%%%%%%%%%%%%%%%%%%%%%%%%%%%%%%%%%%%%%%%%%%%%%%%%%%%
For the purpose, we estimate
$Q_1I_0\bar R_{\epsilon}^2I_0Q_1$.
The annihilation operator with respect to $g^1\oplus g^2\in{\mathfrak h}\oplus{\mathfrak h}$ is represented by operator valued distribution
$a_k^1,a_k^2$ as
\begin{align*}
a(g^1\oplus g^2)=\int dk \left(\bar g^1(k)a_k^1+\bar g^2(k)a_k^2\right).
\end{align*}
Here, $a_k^i,a_k^{i*}$ satisfies $\{a_k^i,a_p^{j*}\}=\delta_{ij}\delta(k-p)$.
By the pull-through formula, 
\begin{align*}
a_k^1L_f=(L_f+\omega(k))a_k^1,\quad
a_k^2L_f=(L_f-\omega(k))a_k^2,
\end{align*}
we get
\begin{align*}
a(g^1\oplus g^2)R_\epsilon^2(e)
=\int dk
\left(
\bar g^1(k)R_\epsilon^2\left(e-\omega(k)\right)a_k^1+
\bar g^2(k)R_\epsilon^2\left(e+\omega(k)\right)a_k^2
\right).
\end{align*}
Using this relation, we obtain the following bound;
\begin{prop}
Under Assumption \ref{O} we have
\[
Q_1I_0\bar R_{\epsilon}^2I_0Q_1
\ge\frac{\pi}{\epsilon}
Q_1\left[\Gamma(e)\otimes 1-C_3\epsilon^{\frac14}\right]Q_1.
\]
Here $C_3$ is a positive constant
which is independent of $\beta_+-\beta_-$ and $v$.
\label{IRIbound}
\end{prop}
Assume that $\theta,\lambda,\epsilon$ satisfies the followings.
\begin{assum}
\begin{align*}
\lambda+\theta\lambda^2\epsilon^{-2}<t_1,\quad
\theta<t_2,\quad
\theta\lambda^2\epsilon^{-1}<t_3,\quad
\epsilon<t_4,\quad
\theta^{-1}\epsilon+\theta\lambda^2\epsilon^{-3}<t_5.
\end{align*}
where
\begin{align*}
t_2\equiv\frac{1}{C_2},\quad
t_3\equiv\frac{v}{8\pi\gamma_e},\quad
t_4\equiv(\frac{\gamma_e}{4C_3})^4,\quad
t_5\equiv\frac{\pi\gamma_e}{4C_2}.
\end{align*}
\label{parameter}
\end{assum}
The second assumption implies
$2\theta\lambda^2-C_2\theta^2\lambda^2
>\theta\lambda^2.$
Using this and the lower bound of $I_0\bar R_{\epsilon}^2I_0$
of Proposition \ref{IRIbound}, we obtain
\begin{multline}
D
\ge\theta\lambda^2
\frac{\pi}{\epsilon}
Q_1\left[\Gamma(e)\otimes 1-C_3(\epsilon^{\frac14})\right]Q_1
-\delta_e\cdot Q_1{\tilde P}_e^{\perp}Q_1
-C_2\left(\lambda^2+\theta^2\lambda^4\epsilon^{-4}\right)Q_1.
\label{Dbound}
\end{multline}
Recall that
$\Gamma(e)$ is bounded below as
$\Gamma(e)\ge \gamma_e\cdot (P(\Gamma(e)=0))^{\perp}$.
Substituting this to (\ref{Dbound}), we get
\begin{multline}
D\ge
\theta\lambda^2
\frac{\pi}{\epsilon}
Q_1\left[\left(\gamma_e-\delta_e\frac{\epsilon}{\pi\theta\lambda^2}
\right)\cdot {\tilde P}_e^{\perp}
-C_3(\epsilon^{\frac14})
-\frac{C_2}{\pi}\left(\theta^{-1}\epsilon
+\theta\lambda^2\epsilon^{-3}\right)
\right]Q_1.
\end{multline}
Now we determine $\delta_e$ as
\begin{align*}
\delta_e=
\frac{\pi\theta\lambda^2}{\epsilon}
\left(\gamma_e+a\right)\quad {\rm with}\quad
0<a<\gamma_e.
\end{align*}
Then we have
\begin{align*}
D\ge
\theta\lambda^2
\frac{\pi}{\epsilon}\left[
-a-C_3(\epsilon^{\frac14})
-\frac{C_2}{\pi}\left(\theta^{-1}\epsilon
+\theta\lambda^2\epsilon^{-3}\right)
\right]
Q_1.
\end{align*}
Hence $\inf\sigma(D)$ is bounded below by
\begin{align*}
\inf\sigma(D)\ge
\theta\lambda^2
\frac{\pi}{\epsilon}\left[
-a-C_3(\epsilon^{\frac14})
-\frac{C_2}{\pi}\left(\theta^{-1}\epsilon
+\theta\lambda^2\epsilon^{-3}\right)
\right]Q_1.
\end{align*}
Recall the lower bound (\ref{Bd}) of $B'$.
As the third condition of Assumption \ref{parameter}
implies
\begin{align*}
\frac 12v-\delta_e
\ge\frac 14v>0,
\end{align*}
and
\begin{align*}
0\ge\theta\lambda^2
\frac{\pi}{\epsilon}\left[
-a-C_3(\epsilon^{\frac14})
-\frac{C_2}{\pi}\left(\theta^{-1}\epsilon
+\theta\lambda^2\epsilon^{-3}\right)
\right],
\end{align*}
the lower bound is
\begin{align*}
B'\vert_{(Q_1\oplus Q_2)\cal H}\ge
\theta\lambda^2
\frac{\pi}{\epsilon}\left[
-a-C_3(\epsilon^{\frac14})
-\frac{C_2}{\pi}\left(\theta^{-1}\epsilon
+\theta\lambda^2\epsilon^{-3}\right)
\right].
\end{align*}
Substituting $[L,A]=B'+\delta_e\tilde{P}_e^{\perp}$, 
and the last two conditions of Assumption \ref{parameter},
we get
\begin{multline*}
PE_\Delta^0[L,A]E_\Delta^0P
\ge\theta\lambda^2
\frac{\pi}{\epsilon}
PE_\Delta^0
\left[
-a
-C_3(\epsilon^{\frac14})
-\frac{C_2}{\pi}\left(\theta^{-1}\epsilon
+\theta\lambda^2\epsilon^{-3}\right)
+\left(\gamma_e+a\right)\cdot
{\tilde P}_e^{\perp}
\right]
E_\Delta^0P\\
=
\theta\lambda^2
\frac{\pi}{\epsilon}
PE_\Delta^0
\left[
\gamma_e
-C_3(\epsilon^{\frac14})
-\frac{C_2}{\pi}\left(\theta^{-1}\epsilon
+\theta\lambda^2\epsilon^{-3}\right)
-\left(\gamma_e+a\right)\cdot
{\tilde P}_e
\right]
E_\Delta^0P\\
\ge
\theta\lambda^2\epsilon^{-1}
PE_\Delta^0\gamma_e
\left[1-7{\tilde P}_e\right]
E_\Delta^0P.
\end{multline*}
Hence we obtain Theorem \ref{L0bound}.$\square$
%%%%%%%%%%%%%%%%%%%%%%%%%%%%%%%%%%%%%%%%%%%%%%%%%%%%%%%%%%%%%%%%%
\subsection{Positive Commutator with respect to
the spectral localization in $L$}
In this subsection we complete the proof of Theorem \ref{positivecommutator}.\\
Let $\Delta'$ be an interval s.t.
$\Delta\subset\Delta'$, $\Delta'\cap\sigma(L_S)=\{e\}$
and ${\rm supp}\varsigma\subset \Delta'$.
Let $F_{\Delta'}\in C_0^\infty({\mathbb R})$ be a smooth
function $0\le F_{\Delta'}\le 1$
which satisfies $F_{\Delta'}=1$ on ${\rm supp}\;\varsigma$ and
${\rm supp}\;F_{\Delta'}\subset \Delta'$.
We set $F_{\Delta'}^0\equiv F_{\Delta'}(L_0)$,
and $\bar F_{\Delta'}^0\equiv 1-F_{\Delta'}^0$.
We evaluate each term of the following equation;
\begin{align}
P\varsigma\left[L,A\right]\varsigma P
&=P\varsigma F_{\Delta'}^0[L,A]F_{\Delta'}^0\varsigma P\label{iti}\\
&+P\varsigma{\bar F_{\Delta'}^0}[L,A]F_{\Delta'}^0\varsigma P
+h.c.\label{ni}\\
&+P\varsigma{\bar F_{\Delta'}^0}[L,A]\bar F_{\Delta'}^0\varsigma P
\label{san}
\end{align}
For the first part (\ref{iti}), we have the following lemma.
\begin{lem}\label{liti}
Suppose that $\lambda,\epsilon,\theta$ satisfy
Assumption \ref{parameter}.
Then we have
\[
(\ref{iti})=P\varsigma F_{\Delta'}^0[L,A]F_{\Delta'}^0\varsigma P
\ge
\frac{\theta\lambda^2}{\epsilon}
\gamma_e P\varsigma(1-\lambda c_1-7\tilde P_e)\varsigma P,
\]
where $c_1$ is a constant which
depends only on $F_{\Delta'}$ and $\Vert I_0\Vert$.
\end{lem}
{\it Proof}\\
The proof is the same as that of 
inequality (63) in \cite{Merk01}.
$\square$\\
Next we evaluate the second part (\ref{ni}).
\begin{lem}\label{lni}
\[
(\ref{ni})=P\varsigma\bar F_{\Delta'}^0[L,A]F_{\Delta'}^0\varsigma P
+h.c.
\ge
-c_2\frac{\theta\lambda^2}{\epsilon}
(\epsilon\theta^{-1}+\epsilon+\lambda\epsilon^{-1})
P\varsigma^2P
\]
where $c_2$ is a constant
which depends only on $F_{\Delta'}$ and $I_0$.
\end{lem}
{\it Proof}\\
We divide (\ref{ni}) into three parts;
\begin{align}
(\ref{ni})
&=P\varsigma\bar F_{\Delta'}^0S_1F_{\Delta'}^0\varsigma P
+h.c.
\label{niiti}\\
&+\lambda P\varsigma\bar F_{\Delta'}^0I_1F_{\Delta'}^0\varsigma P
+h.c.
\label{nini}\\
&+P\varsigma\bar F_{\Delta'}^0[L,b]F_{\Delta'}^0\varsigma P.
+h.c.
\label{nisan}
\end{align}
Using the fact that $0\le F_{\Delta'}^0\le 1$, we get
$(\ref{niiti})\ge 0$ as in the case (57) of \cite{Merk01}.

To evaluate (\ref{nini}), we note 
$\varsigma\bar F_{\Delta'}^0=\varsigma(L)(\bar F_{\Delta'}(L_0)-\bar F_{\Delta'}(L))$.
By operator calculous, we have the estimation
$\Vert \varsigma\bar F_{\Delta'}^0\Vert\le c_1''\vert \lambda\vert$.
As $I_1$ is bounded, we obtain 
$(\ref{nini})\ge -c_1'\lambda^2$.
Here, $c_1'',c_1'$ depends only on $I_0$ and $\bar F_{\Delta'}^0$.

The last part (\ref{nisan}) can be estimated in a manner 
of Proposition 5.2 in \cite{Merk01}:
\begin{multline*}
(\ref{nisan})
=P\varsigma\bar F_{\Delta'}^0[L,b]F_{\Delta'}^0\varsigma P+h.c.
=
P\varsigma
\left(\sqrt{\bar F_{\Delta'}(L_0)}-\sqrt{\bar F_{\Delta'}(L)}\right)
\sqrt{\bar F_{\Delta'}^0}[L,b]
F_{\Delta'}^0\varsigma P+h.c.\\
\ge
-c_2^{''}\left(\theta\lambda^2+\theta\lambda^3\epsilon^{-2}\right)\varsigma^2P
=-c_2^{''}\frac{\theta\lambda^2}{\epsilon}
\left(\epsilon+\lambda\epsilon^{-1}\right)\varsigma^2P.
\end{multline*}
Hence we have 
\begin{align*}
(\ref{ni})=(\ref{niiti})+(\ref{nini})+(\ref{nisan})
\ge
-c_2\frac{\theta\lambda^2}{\epsilon}
\left(\theta^{-1}\epsilon+\epsilon+\lambda\epsilon^{-1}\right)\varsigma^2P,
\end{align*}
where $c_2$ depends only on $I_0$ and
$F_{\Delta'}$.$\square$

Finally, the third part (\ref{san}) can be estimated as follows.
\begin{lem}\label{lsan}
\[
P\varsigma{\bar F_{\Delta'}^0}[L,A]\bar F_{\Delta'}^0\varsigma P
\ge
-c_3\frac{\theta\lambda^2}{\epsilon}
(\theta^{-1}\epsilon+\lambda\epsilon^{-1})\varsigma^2P.
\]
where $c_3$ depends only on $I_0$ and $F_{\Delta'}$.
\end{lem}
{\it Proof}\\
The same as that of Proposition 5.2 in \cite{Merk01}.
$\square$\\
Now let us complete the proof of Theorem \ref{positivecommutator}.
By Lemma \ref{liti} to \ref{lsan}, 
if $\lambda,\epsilon$ and $\theta$ satisfy Assumption \ref{parameter},
we obtain
\begin{align*}
P\varsigma\left[L,A\right]\varsigma P=(\ref{iti})+(\ref{ni})+(\ref{san})
\ge
\frac{\theta\lambda^2}{\epsilon}\gamma_e
P\varsigma\left(
1-7\tilde P_e
-\left(\lambda C_4+\frac{C_5}{\gamma_e}(
\epsilon\theta^{-1}+\epsilon+\lambda\epsilon^{-1})\right)
\right)\varsigma P,
\end{align*}
where $C_4,C_5$ depends only on $F_\Delta'$ and $I_0$. 
Let $\epsilon$, $\theta$ be
\begin{align*}
\epsilon=\lambda^{\frac{44}{100}},\quad
\theta=\lambda^{\frac{26}{100}}.
\end{align*}
Then 
\begin{align*}
\epsilon\theta^{-1}=\lambda^{\frac{18}{100}},\quad
\lambda\epsilon^{-1}=\lambda^{\frac{56}{100}},\quad
\theta\lambda^2\epsilon^{-1}=\lambda^{\frac{182}{100}},\quad
\theta\lambda^2\epsilon^{-2}
=\lambda^{\frac{138}{100}},\quad
\theta\lambda^2\epsilon^{-3}=\lambda^{\frac{94}{100}}
\end{align*}
and if $\lambda$ is sufficiently small, Assumption \ref{parameter}
is satisfied.
Further more, for $\lambda$ small enough,
\begin{align}
\lambda C_4+\frac{C_5}{\gamma_e}\left(
\epsilon\theta^{-1}+\epsilon+\lambda\epsilon^{-1})\right)
<\frac{1}{2}.
\label{ab}
\end{align}
is satisfied.
Hence we obtain
\begin{align*}
P\varsigma\left[L,A\right]\varsigma P
\ge
\frac{\theta\lambda^2}{\epsilon}
\frac{\gamma_e}{2}P\varsigma\left(1-14\tilde P_e\right)\varsigma P
=\lambda^{\frac{182}{100}}
\frac{\gamma_e}{2}P\varsigma\left(1-14\tilde P_e\right)\varsigma P.
\end{align*}
Now let us estimate the range of $\lambda$,
which satisfies the Assumption \ref{parameter} and
(\ref{ab}).
We assume $\beta_0<\beta_+,\beta_-<\beta_1$ for any fixed 
$0<\beta_0<\beta_1<\infty$,
and the bound of the form factor $f\in{\mathfrak h}$
$\Vert p f\Vert,\Vert f\Vert\le b$ for fixed $0<b<\infty$.
Under these bounds, one can easily check that
there exists a constant $C$ which is independent of $v$ and $\beta_+-\beta_-$,
such that if
$\lambda$ satisfies 
\begin{align*} 
\lambda<C
\min\{
v^{\frac{100}{26}},\left(\frac{v}{\gamma_e}\right)^{\frac{100}{182}},
(\gamma_e)^{\frac{100}{11}},(v\gamma_e)^{\frac{100}{18}}
\},
\end{align*}
then the Assumption \ref{parameter} and
(\ref{ab}) are satisfied.
Let 
\begin{align*}
\lambda_1=
C
\min\left\{
v^{\frac{100}{26}},\left(\frac{v}{\gamma_e}\right)^{\frac{100}{182}},
(\gamma_e)^{\frac{100}{11}},(v\gamma_e)^{\frac{100}{18}}
\right\},
\end{align*}
and we obtain Theorem \ref{positivecommutator}.$\square$\\
%%%%%%%%%%%%%%%%%%%%%%%%%%%%%%%%%%%

\section{Virial Theorem}\label{Virial}
In this section, we complete the proof of  Theorem \ref{spectral}.
We apply the new method introduced by M.Merkli \cite{Merk01}
to treat the domain question.
%M.Merkli introduced $C^{\infty}$-approximation function $f_\alpha$
%and $g_\nu$ to treat the problem for finite temperature case \cite{Merk01}.
He solved the problem by approximating the eigenvector
of $L$ by vectors in the domain of $N$ and $A_0$.

Assume that $\psi$ is a normalized eigenvector
of $L$ with eigenvalue $e$.
Let $f$ be a bounded $C^{\infty}$-function such that $f'\ge 0$,
$f'(0)=1$ and let $g$ be a bounded $C_0^{\infty}$-function
with support in the interval $[-1,1]$.
Define the operators $f_\alpha\equiv f(i\alpha A_0)$,
$h_\alpha\equiv\sqrt{f'(i\alpha A_0)}$ and $g_\nu\equiv(\nu N)$.
When $\alpha,\nu$ goes to zero,
$h_\alpha,g_\nu$ strongly converges to $1$,.
By approximating the eigenvector $\psi$ by
$h_\alpha g_\nu\psi$, we can carry out the arguments regoliously. 

As the proof goes parallel to \cite{Merk01},
we just comment on the differences.
We define $\tilde e(e)$ as in Theorem \ref{spectral}.
For simplicity, we use the notations $K\equiv [L,A_0]=S_1+I_1^\lambda$,
$\langle A\rangle_{\psi}\equiv\langle\psi, A\psi\rangle$,
$\psi_{\alpha,\nu}\equiv h_\alpha g_\nu\psi$, and 
$I_j^\lambda\equiv \lambda I_j$.
The proof is done by evaluating 
the upper and the lower bound of
$\left\langle K\right\rangle_{\psi_{\alpha,\nu}}
=\left\langle \psi_{\alpha,\nu}, K\psi_{\alpha,\nu}\right\rangle$.

The estimation of the upper bound is done by the 
expansion of commutators, using operator calculous:
\begin{align}
&\left\langle
K\right\rangle_{\psi_{\alpha\nu}}
=
-\alpha^{-1}\nu{\rm Im}\left\langle\psi
\vert g_\nu f_\alpha \tilde I_0 \psi\right\rangle
+\alpha^2\left(
\left\langle\psi\vert g_\nu R
g_\nu\psi\right\rangle+{\rm Re}\left\langle\psi\vert g_\nu R'
g_\nu\psi\right\rangle
\right).
\end{align}
Here, $\tilde I_0$, $R$ and $R'$ are given by
\begin{align*}
&\tilde I_0
=2\int d\tilde g(z)(z-\nu N)^{-1}ad_N^1(I_0^\lambda)(z-\nu N)^{-1},\\
&R=-\frac 14 \int d\tilde f^{''}(z)
(z-i\alpha A_0)^{-1}ad_{A_0}^3(L_0+I_0^\lambda) (z-i\alpha A_0)^{-1}\\
&-\frac12
\int d\tilde h(z)
(z-i\alpha A_0)^{-2}\left[h_\alpha,ad_{A_0}^3(L_0+I_0^\lambda)\right] (z-i\alpha A_0)^{-1}\\
&+\frac12\int d\tilde h(z)
(z-i\alpha A_0)^{-1}h'_\alpha ad_{A_0}^3(L_0+I_0^\lambda)
(z-i\alpha A_0)^{-1}\\
&R'=
\int d\tilde f(z)
(z-i\alpha A_0)^{-3}ad_{A_0}^3(L_0+I_0^\lambda)(z-i\alpha A_0)^{-1}
\end{align*}
where $ad_{M}^k$ is a $k-$fold commutator 
$[\cdots[\cdot,M],M,\cdots,M]$ with $M$.
In the case of \cite{Merk01}, 
the commutators are given by
\begin{align*}
ad_{A_0}^1(L_0)=N,
\quad ad_{A_0}^2(L_0)=ad_{A_0}^3(L_0)=0,
\end{align*}
and $ad_{A_0}^k(I_0^\lambda)$ and $ad_N^1(I_0^\lambda)$
are $N^{\frac12}$-bounded.
Hence $\left\langle K\right\rangle_{\psi_{\alpha\nu}}$ is estimated as
\begin{align}
\left\langle
K\right\rangle_{\psi_{\alpha\nu}}
\sim O(\alpha^2\nu^{-\frac 12}+\nu^{\frac 12}\alpha^{-1}).
\end{align}
In our case, the $k$-fold commutators don't vanish:
\begin{align*}
ad_{A_0}^1(L_0)=S_1,
\quad ad_{A_0}^2(L_0)=S_2,\quad ad_{A_0}^3(L_0)=S_3.
\end{align*}
On the other hand,
the interaction terms $ad_{A_0}^k(I_0)$, and $ad_N^1(I_0^\lambda)$
are bounded.
Hence $\left\langle K\right\rangle_{\psi_{\alpha\nu}}$ is estimated as
\begin{align}
\left\langle
K\right\rangle_{\psi_{\alpha\nu}}
\sim O(\alpha^2\nu^{-1}+\alpha^{-1}\nu).
\end{align}
The difference emerges from two factors:
the non-commutativity of $S_i$ with $A_0$,
and the boundedness of the interaction.
The first one makes things worse,
while the second one makes it easier.

The estimation of the lower bound also goes parallel to \cite{Merk01}.
As the positive commutator is localized with respect to 
the spectrum of $L$,
we need to decompose the Hilbert space.
Let $\Delta$ be an interval which contains $e$.
Suppose that $\Delta$ contains 
exactly one eigenvalue $\tilde e(e)$ of $L_S$
i.e., $\Delta\cap \sigma(L_S)=\{\tilde e(e)\}$,
and $\tilde e(e)$ belongs to the interior of $\Delta$.
In \cite{Merk01}, M.Merkli introduced a partition of unity
\begin{align*}
\chi_\Delta^2+\bar\chi_\Delta^2=1.
\end{align*}
Here $\chi_\Delta\in C_0^\infty({\mathbb R})$
is a smooth function s.t.$\chi_\Delta=1$ on $\Delta$
and ${\rm supp}\; \chi_\Delta\cap\sigma(L_S)=\{\tilde e(e)\}$.
By Theorem \ref{positivecommutator}, we have 
\begin{align}
P\chi_\Delta(L)\left[L,A\right]\chi_\Delta(L)P
\ge
b_{{\tilde e}(e)}
P\chi_\Delta(L)\left(1-14\tilde P_{{\tilde e}(e)}\right)\chi_\Delta(L)P,
\label{final}
\end{align}
for $0<\vert \lambda\vert <\lambda_1$,
where $b_e=\frac{1}{2}\cdot\lambda^{\frac{182}{100}}\cdot\gamma_e$.

Another partition of unity is also needed:
\begin{align*}
\chi^2+\bar\chi^2=1,
\end{align*}
where $\chi\in C^\infty$ satisfies
$\chi(t)=1$ for $\vert t\vert\le \frac 12$ and
$\chi(t)=0$ for $1\le \vert t\vert$.
And he set $\chi_n=\chi(N/n),\bar\chi_n^2=1-\chi_n^2$
for $0<n<1/\nu$.
With respect to the partition of unity, he obtained the lower bound
\begin{align*}
\langle K\rangle_{\psi_{\alpha,\nu}}
\ge
b_{{\tilde e}(e)}\Vert \psi_{\alpha,\nu}\Vert^2
-Cb_{{\tilde e}(e)}\delta_{{{\tilde e}(e)},0}\Vert \tilde P_{0}\chi_\Delta\chi_n\psi_{\alpha,\nu}\Vert^2\\
-O(\epsilon\eta^{-1}+\eta+\alpha n^{\frac32}+n^{-\frac12})
-Cb_{{\tilde e}(e)}(n^{-1}+n^{-3}+\alpha^2n^2),
\end{align*}
where $C$ is a positive constant and
$\eta,\epsilon>0$ are arbitrary positive parameter
which satisfy
$\frac{n}{2}-C\eta^{-1}\epsilon^{-2}\ge b_{{\tilde e}(e)}$.
Here we represented the result in our notations.

The argument can be carried out parallel in our case.
We just need to take care of the projection $P=P(N_{\Lambda_v^c}=0)$, because
the positive commutator is localized to the range of 
it.
This is easily done by the strong commutativity of 
$N_{\Lambda_v^c}$ with and $K$ and $N$:
\begin{align*}
\left\langle K\right\rangle_{\psi_{\alpha,\nu}}
\ge\left\langle K\right\rangle_{P\psi_{\alpha,\nu}}
-\Vert I_1^\lambda\Vert\cdot \Vert\bar P\psi_{\alpha,\nu}\Vert^2.
\end{align*}
Here we used $\bar P S_1 \bar P\ge 0$.
Then we obtain
\begin{multline}
O(\alpha^2\nu^{-1}+\alpha^{-1}\nu)\sim\left\langle K\right\rangle_{\psi_{\alpha,\nu}}\\
\ge
b_{{\tilde e}(e)}\left(\Vert\chi_\Delta P\chi_n\psi_{\alpha\nu}\Vert^2
-14\Vert \tilde P_{{\tilde e}(e)}\chi_\Delta P\chi_n\psi_{\alpha\nu}\Vert^2\right)
-\Vert I_1^\lambda\Vert
\Vert P\bar\chi_\Delta\chi_n \psi_{\alpha,\nu}\Vert^2\\
-C_3(n^{-1}+\nu+\alpha\cdot n+\alpha)-\Vert I_1^\lambda\Vert\cdot\Vert\bar P\psi_{\alpha,\nu}\Vert^2
-\Vert I_1^\lambda\Vert\cdot\Vert\bar\chi_n\psi_{\alpha,\nu}\Vert^2
-n^{-2}\cdot C_2.
\label{lbb}
\end{multline}
Substituting $\nu=\alpha^{\frac 32}$ and $n=\alpha^{-\frac 12}$
and taking $\alpha\to 0$ limit, we obtain
\begin{align}
0=\lim_{\alpha\to 0}\left\langle K\right\rangle_{\psi_{\alpha,\nu}}
\ge
b_{{\tilde e}(e)}\left(\Vert\psi\Vert^2
-14\Vert \tilde P_{{\tilde e}(e)}\psi\Vert^2\right).
\label{lim}
\end{align}
If $\tilde P_{{\tilde e}(e)}\psi=0$, the inequality 
(\ref{lim}) is a contradiction.
Hence for $0<\vert\lambda\vert<\lambda_1$,
there is no eigenvector of $L$ with eigenvalue $e$
which is orthogonal to $\tilde P_{{\tilde e}(e)}$.
Recalling that $\lambda_1$ can be taken as
(\ref{lam1}), we obtain Theorem \ref{spectral}.

One might wonder if $\chi_n$ is necessary in our case
where the interaction term is bounded.
Note that the right hand side of (\ref{lbb}) has a term of order $\alpha n$,
while the left hand side has a term of order $\alpha^{-1}\nu$.
Without $\chi_n$, $\alpha n$ is replaced by $\alpha\nu^{-1}$.
We can't make $\alpha\nu^{-1}$ and $\alpha^{-1}\nu$
converge to zero simultaneously, in any choice of $\nu$.
So $\chi_n$ is still required.
%%%%%%%%%%%%%%%%%%%%%%%%%%%%%%%%%%%%%%%%%%%%%%%%%%%
%%%%%%%%%%%%%%%%%%%%%%%%%%%%%%%%%%%%%%%%%%%%%%%%%%%%%%%%%%%%%%%%%%
\section{The Stability of the NESS}\label{stability}
In this section we investigate our physical interest: the stability
of the NESS.
By proving Theorem \ref{gamma}, we complete the proof of
Theorem \ref{unstable} and \ref{frte}.
Recall that the NESS of the free Fermion model is given by 
$n$-point functions (\ref{Fdis})
with a distribution function (\ref{rho}).
We show if the NESS is 
far from equilibrium,
i.e., the inverse temperature $\beta_-$ and $\beta_+$
are different,
it is macroscopically unstable (Theorem \ref{unstable}).
This results is due to the following fact:
for the NESS far from equilibrium,
the number of the particles with momentum $k$
is different from the number of the particles with momentum $-k$,
although they have the same energy.
On the other hand, for a class of interaction,
we show {\it return to equilibrium} (Theorem \ref{frte}).
%%%%%%%%%%%%%%%%%%%%%%%%%%%%%%%%%%%%%%%%%%%%%%%%%%%%%%%%%%%%%%%%%
\subsection{\it Instability of the NESS}
In this subsection, we investigate the instability of NESS
under the interaction with small system.
For the purpose, we study the kernel of $\Gamma(e)$.
Recall the definition of 
$\varphi_n,E_n,E_{nm},N_l^{(j)},N_r^{(i)},N_l,N_r$,
given in Assumption \ref{small2}.
As $H_S$ is finite dimensional, $N_r^c\neq\varnothing$
and $N_l^c\neq\varnothing$ for $e\neq 0$.\\
By a straight forward calculation,
we obtain
\begin{align}
\Gamma(e)
=\int_{-\pi}^\pi dk \;\sum_{E_{n,m}\neq e}
F_{n,m}^{1*}(k)F_{n,m}^1(k)
\delta\left(\omega\left(k\right)+E_{n,m}-e\right)\nonumber\\
+F_{n,m}^{2*}(k)F_{n,m}^2(k)
\delta\left(-\omega\left(k\right)+E_{n,m}-e\right)
\label{gamsimple}
\end{align}
where
\begin{align*}
F_{n,m}^{i}(k)
=p_nYp_{N_l^{(m)}}\otimes p_m  g_1^i(k)
-p_n\otimes p_m\bar Yp_{N_r^{(n)}}g_2^i(k).
\end{align*}
First let us consider $e\neq 0$ case.
Note that each term of (\ref{gamsimple}) is positive.
So we take the sum over the following subset
\begin{align*}
(n,m)\in
N_l\times N_r^c\;{\dot \cup}\;N_l^c\times N_r.
\end{align*}
and ignore the other contributions to estimate the lower bound.
Note that $N_l,N_r,N_l^c$ and $N_r^c$ are all non-empty in case
$e\neq 0$.
We have the following relations:
if $n\in N_l$, then $N_r^{(n)}\neq\varnothing$,
if $m\in N_r$, then $N_l^{(m)}\neq\varnothing$,
if $n\in N_l^c$, then $N_r^{(n)}=\varnothing$, and
if $m\in N_r^c$, then $N_l^{(m)}=\varnothing$.
Especially, $E_{n,m}\neq e$ for
$(n,m)\in N_l\times N_r^c\;{\dot \cup}\;N_l^c\times N_r$.
Hence for all $(n,m)\in N_l\times N_r^c$, we have
\begin{align*}
F_{n,m}^{i}(k)
=-p_n\otimes p_m\bar Yp_{N_r^{(n)}}g_2^i(k),
\end{align*}
and
for $(n,m)\in N_l^c\times N_r$, we have
\begin{align*}
F_{n,m}^{i}(k)
=p_nYp_{N_l^{(m)}}\otimes p_m  g_1^i(k).
\end{align*}
Then under Assumption \ref{nondegenerate} and \ref{small2}, we have
\begin{multline*}
\Gamma(e)
\ge
b_0\left(
\sum_{n\in N_l}
p_n\otimes p_{N_r^{(n)}}\bar Yp_{N_r^c}\bar Yp_{N_r^{(n)}}
+\sum_{m\in N_r}
p_{N_l^{(m)}}Yp_{N_l^c}Yp_{N_l^{(m)}}\otimes p_m 
\right)\\
\ge
b_0\cdot P(L_S=e)\delta_0=\gamma_e\cdot P(L_S=e),
\end{multline*}
with some $b_0>0$ and $\gamma_e=b_0\delta_0>0$.
Here, $b_0$ is a constant which is independent of 
$\beta_+,\beta_-$ in the interval
$(\beta_0,\beta_1)$ for fixed $0<\beta_0<\beta_1<\infty$.
Hence for $e\neq 0$, 
we have ${\rm Ker}\Gamma(e)=\{0\}$, and $\tilde{P}_e=0$.

Next let us consider $e=0$ case.
In this case, we can not apply the above arguments
because $N_l^c=N_r^c=\varnothing$.
By the first condition of Assumption \ref{nondegenerate},
a vector in $P(L_S=0)$ is of the form
\begin{align*}
\varphi=\sum_{i}c_i\varphi_i\otimes\varphi_i.
\end{align*}
Note that Assumption \ref{nondegenerate} imply $N_l^{(n)}=N_r^{(n)}=\{n\}$.
We have
\begin{multline}
\left\langle\varphi\left\vert\Gamma(0)
\right\vert\varphi\right\rangle
=\sum_{n\neq m}\vert Y_{mn}\vert^2\\
\left[\left\vert c_m\cdot \left(1-\rho\right)^{\frac12}(q_{mn})
-c_n\cdot \rho^{\frac12}(q_{mn})\right\vert^2\right]\vert f(q_{mn})\vert^2
\int_0^\pi dk \delta\left(\omega\left(k\right)+E_{n,m}\right)\\
+\left[\left\vert c_m\cdot \left(1-\rho\right)^{\frac12}(-q_{mn})
-c_n\cdot \rho^{\frac12}(-q_{mn})\right\vert^2\right]\vert f(-q_{mn})\vert^2
\int_{-\pi}^0 dk \delta\left(\omega\left(k\right)+E_{n,m}\right)\\
+
\left[
\left\vert c_m\cdot \rho^{\frac12}(q_{nm})
-c_n\cdot \left(1-\rho\right)^{\frac12}(q_{nm})\right\vert^2
\right]\vert f(q_{nm})\vert^2
\int_0^\pi dk \delta\left(-\omega\left(k\right)+E_{n,m}\right)\\
+\left[
\left\vert c_m\cdot \rho^{\frac12}(-q_{nm})
-c_n\cdot \left(1-\rho\right)^{\frac12}(-q_{nm})\right\vert^2
\right]\vert f(-q_{nm})\vert^2
\int_{-\pi}^0 dk\delta\left(-\omega\left(k\right)+E_{n,m}\right),
\label{gam0}
\end{multline}
where $\omega(q_{mn})=E_{mn}$.
By Assumption \ref{O} and Assumption \ref{nondegenerate},
$f(q_{mn})$, $f(-q_{mn})$, $f(q_{nm})$, $f(-q_{nm})$ are non-zero,
and the integrals including $\delta$-function give strictly
positive contributions.
So there exist strictly positive $a_0$,
such that
\begin{align}
(\ref{gam0})\ge \sum_{n\neq m}a_0
\left\{\left\vert c_m\cdot \left(1-\rho\right)^{\frac12}(q_{mn})
-c_n\cdot \rho^{\frac12}(q_{mn})\right\vert^2
+\left\vert c_m\cdot \left(1-\rho\right)^{\frac12}(-q_{mn})
-c_n\cdot \rho^{\frac12}(-q_{mn})\right\vert^2\right.\nonumber\\
\left.+\left\vert c_m\cdot \rho^{\frac12}(q_{nm})
-c_n\cdot \left(1-\rho\right)^{\frac12}(q_{nm})\right\vert^2
+\left\vert c_m\cdot \rho^{\frac12}(-q_{nm})
-c_n\cdot \left(1-\rho\right)^{\frac12}(-q_{nm})\right\vert^2
\right\}.
\label{gam1}
\end{align}
Hence the necessary condition for ${\rm Ker}\Gamma(0)$ to be
non-trivial is the existence of $\{c_n\}$ which satisfies 
\begin{align*}
\left\vert c_m\cdot \left(1-\rho\right)^{\frac12}(q_{nm})
-c_n\cdot \rho^{\frac12}(q_{mn})\right\vert^2
=\left\vert c_m\cdot \left(1-\rho\right)^{\frac12}(-q_{mn})
-c_n\cdot \rho^{\frac12}(-q_{mn})\right\vert^2\\
=\left\vert c_m\cdot \rho^{\frac12}(q_{nm})
-c_n\cdot \left(1-\rho\right)^{\frac12}(q_{nm})\right\vert^2
=\left\vert c_m\cdot \rho^{\frac12}(-q_{nm})
-c_n\cdot \left(1-\rho\right)^{\frac12}(-q_{nm})\right\vert^2
=0.
\end{align*}
This implies the following:
\begin{align}
\frac{c_n}{c_m}
=\frac{\left(1-\rho\right)^{\frac12}(q_{mn})}{\rho^{\frac12}(q_{mn})}
=\frac{\left(1-\rho\right)^{\frac12}(-q_{mn})}{\rho^{\frac12}(-q_{mn})}
=\frac{\rho^{\frac12}(q_{nm})}{\left(1-\rho\right)^{\frac12}(q_{nm})}
=\frac{\rho^{\frac12}(-q_{nm})}{\left(1-\rho\right)^{\frac12}(-q_{nm})}
\label{detail}
\end{align}
for $n\neq m$.
Recall that 
the Fermion distribution
$\rho$ in the NESS is given by
\begin{align*}
\rho(k)\equiv\left \{
\begin{gathered}
\left(1+e^{\beta_{-}
\left( \cos\left(k \right)-\gamma \right)}\right)^{-1},\quad
k\in [0,\pi)\\
\left(1+e^{\beta_{+}
\left( \cos\left(k \right)-\gamma \right)}\right)^{-1},\quad
k\in [-\pi,0).
\end{gathered}
\right.
\end{align*}
The condition (\ref{detail}) requires
\begin{align*}
e^{\beta_{-}E_{nm}}=e^{\beta_{+}E_{nm}},
\end{align*}
i.e., $\beta_-=\beta_+$.
That is, $\Gamma(0)$ has non trivial kernel
only if the NESS is an equilibrium state indeed.
Otherwise, we have $\Gamma(0)\ge\gamma_0\cdot 1>0$.
So, if $\beta_-\neq\beta_+$, we have $\tilde P_e=0$ for
all eigenvalue $e$ of $L_S$
and obtain the first statement of Theorem \ref{gamma}.

Now let us complete the proof of Theorem \ref{unstable}.
Combining Theorem \ref{spectral} and \ref{gamma},
the Liouville operator $L$
corresponding to the NESS does not have any eigenvector
for $0<\vert \lambda\vert<\lambda_1$.
Hence the NESS is macroscopically unstable by Proposition \ref{sigular}.
We fix the bound $\beta_0<\beta_+,\beta_-<\beta_1$,
$\Vert p f\Vert,\Vert f\Vert\le b$ for any fixed 
$0<\beta_0<\beta_1<\infty$ and $0<b<\infty$.
Let us estimate the dependence of $\lambda_1$ (\ref{lam1})
on $\beta_+-\beta_-$ for fixed $v$.
For $e\neq 0$, we have $\gamma_e= b_0\delta_0$ as seen in above,
which is independent of $\beta_+-\beta_-$.
On the other hand, $\gamma_0$ converges to $0$ as
$\beta_+-\beta_-$ goes to $0$.
Substituting
\begin{align*}
c_n=\frac{{\rm e}^{-\frac{\beta_-}{2}E_n}}{\sqrt{{Z_{\beta_-}}}},\quad
Z_{\beta_-}=\sum_{n}{\rm e}^{-\beta_-E_n}
\end{align*}
to (\ref{gam1}), we can estimate
$(\ref{gam0})\ge C(\beta_+-\beta_-)^2\ge\gamma_0>0$,
i.e.
$\gamma_0\sim O(\beta_+-\beta_-)^2$.
Hence we have $\lambda_1\sim O\vert\beta_+-\beta_+\vert^{\frac{200}{11}}$.
On the other hand, if we fix $\beta_+$ and $\beta_-$,
we have $\lambda_1(v)\sim v^{\frac{50}{9}}\to 0$
as $v\to 0$.
Then, we obtain Theorem
\ref{unstable}.
%%%%%%%%%%%%%%%%%%%%%%%%%%%%%%%%%%%%%%%%%%%%%%%%%%%%%%%%%%%%%%%
\subsection{\it Return to equilibrium}
In this section we
investigate the equilibrium case, i.e., $\beta_+=\beta_-$.
By the result of the previous subsection, we can show 
{\it return to equilibrium} for the class of interaction we introduced.
The following Theorem was shown by H.Araki
\cite{AraRE1},\cite{AraRE2}:
\begin{thm}\label{PKMS}
Let $({\mathfrak A},\tau)$ be a $C^*$-dynamical system and let
$\omega$ be a $(\beta,\tau)$-KMS state
with GNS-representation $({\cal H},\pi,\Omega)$.
Let $L$ be the Liouville operator corresponding to $\tau$.
If $P=P^*\in{\mathfrak A}$
then $\Omega\in D({\rm e}^{\beta(L+\pi(P))/2})$.
Let $\tau^P$ be the perturbed automorphism group by $P$,
and let $\Omega^P\equiv{\rm e}^{\beta(L+\pi(P))/2}\Omega$.
Then the state $\omega^P$ defined by
\[
\omega^P(A)=\frac{(\Omega^P,A\Omega^P)}{(\Omega^P,\Omega^P)}
\]
is a $(\beta,\tau^P)$-KMS state.
\end{thm}
On the other hand, the following theorem
concerning {\it return to equilibrium} is known
\cite{BFSRE}:
\begin{prop}\label{return}
Let $({\mathfrak A},\tau)$ be a $C^*$-dynamical system and let
$\omega$ be a $(\beta,\tau)$-KMS state
with the GNS-representation $({\cal H},\pi,\Omega)$.
Assume that the Liouville operator $L$ of $\tau$
has a simple eigenvalue $0$ corresponding to the eigenvector $\Omega$,
and that the rest of the spectrum of $L$ is continuous.
Then for any $\omega$-normal state $\eta$,
we have the return to equilibrium in an ergodic mean sense:
\[
\lim_{T\to\infty}\frac{1}{T}\int_0^T\eta\left(\alpha_t(A)\right)dt
=\omega(A),
\quad\quad A\in{\mathfrak A}.
\]
\end{prop}

Let us return to our system.
Now we have $\beta\equiv\beta_-=\beta_+$.
First let us investigate the kernel of $\Gamma(e)$.
As in the previous subsection, ${\rm Ker}(e)=\{0\}$
for $e\neq 0$.
On the other hand, the equation (\ref{gam0}) implies that ${\rm Ker}\Gamma(0)$ 
is one-dimensional and is spanned by a vector of the form
\begin{align*}
\varphi=\sum_{i}c_i\varphi_i\otimes\varphi_i,
\end{align*}
with
\begin{align*}
c_n=\frac{{\rm e}^{-\frac{\beta}{2}E_n}}{\sqrt{{Z_{\beta}}}},\quad
Z_{\beta}=\sum_{n}{\rm e}^{-\beta E_n}.
\end{align*}
Hence we obtain Theorem \ref{gamma}.

Second, note that $\omega_\rho$ is a $(\beta,\alpha_t^f)$-KMS state
of ${\cal O}_f$.
We denote by $\omega_\beta^S$ the $(\beta,\alpha_S)$-KMS state
over ${\cal O}_S$.
The state $\omega_\beta^S\otimes\omega_\rho$ is 
a $(\beta,\alpha_0)$-KMS state over 
${\cal O}_S\otimes{\cal O}_f$.
As our perturbation $V$ is an element of
${\cal O}$, the nontriviarity of 
${\rm Ker}L$ is guaranteed
by Theorem \ref{PKMS}.
The fact that $0$ is the simple eigenvalue of $L$
can be derived as follows :
let $\psi_1,\psi_2$ be eigenvectors of $L$
with eigenvalue $0$.
By Theorem \ref{spectral},
we have $\tilde{{P_0}}\psi_i=c_i\varphi\otimes\Omega$, with $c_i\neq 0$.
Then $\frac{\psi_1}{c_1}-\frac{\psi_2}{c_2}$ is an eigenvector
of $L$ with eigenvalue $0$, which is orthogonal to $\tilde{P_0}$.
By Theorem \ref{spectral}, this entails 
$\psi_1=\frac{c_1}{c_2}\psi_2$.
Hence $0$ is the simple eigenvalue of $L$.
We denote the corresponding eigenvector by $\Omega_V$
and the state corresponding to $\Omega_V$ by $\omega_V$.
By Theorem \ref{PKMS}, $\omega_V$ is a $(\beta,\alpha)$-KMS state.
On the other hand, as the kernel of $\Gamma(e)$ is trivial for $e\neq 0$,
$L$ has no other eigenvalue.
So the rest of the spectrum of $L$ is continuous.
Accordingly, from the Proposition \ref{return},
we obtain the {\it return to equilibrium} Theorem \ref{frte}.\\
%%%%%%%%%%%%%%%%%%%%%%%%%%%%%%%%%%%%%%%%%%%%%%%%%%%%%%%%%%%%%%%%
\noindent
{\bf Acknowledgement.}\\
{The author thanks for A.Arai, F.Hiroshima, and T.Matsui for 
useful advices. Many thanks also go to M.Merkli for helpful arguments and comments.}
\noindent
%%%%%%%%%%%%%%%%%%%%%%%%%%%%%%%%%%%%%%%%%%%%%%%%%%%%%%%%%%%%%%
\appendix
\section{An Example which satisfies Assumptions \ref{suppg} to \ref{small2}}
\label{exam}
We give an example in dimension $d=2$.
We consider $\gamma=0$ case.
We fix some $0<v<1$, and define $0\le k_v\le \pi/2$ by
$\sin^2 k_v=v$.
Let $H_S=b\sigma_z$, with $0<b<(\cos k_v)/4$,
and define $0\le k_{4b}<k_{2b}<\frac{\pi}{2}$ by $4b=\cos k_{4b}$,
$2b=\cos k_{2b}$.
We have then $0<k_v<k_{4b}<k_{2b}\le \frac{\pi}{2}$,
and $S=\{k_{4b},-k_{4b},\pi-k_{4b},-\pi+k_{4b},
k_{2b},-k_{2b},\pi-k_{2b},-\pi+k_{2b}
\}$ are all in $\Lambda_v$.
We choose $f$ as a smooth function with support in $\Lambda_v$,
which takes non-zero values on $S$.
As the interaction, we take $Y$ as
\begin{align*}
Y=
\begin{pmatrix}
0&1\\
1&0
\end{pmatrix}
\end{align*}
Because $p$ acts on $C^\infty$ as (\ref{acp}),
$f$ is in $D(p^3)$ and Assumption \ref{suppg} is satisfied.
The eigenvalue of $L_S$ is $2b,0,-2b$.
We have $\cos(k_{4b})=4b,\cos(k_{2b})=2b,
\cos(\pi-k_{2b})=-2b,\cos(\pi-k_{4b})=-4b$
and as $S$ is included in $\Lambda_v$,
Assumption \ref{O} is satisfied.
Assumption \ref{nondegenerate} is trivial.
Let us check the Assumption \ref{small2}.
Let ${\varphi_0,\varphi_1}$ be eigenvectors of $H_S$,
corresponding to eigenvalue $-b,b$ respectively.
For $e=2b$, $N_l^{(0)}=\{1\}$, $N_l^{(1)}=\varnothing$,
$N_r^{(0)}=\varnothing$ and $N_r^{(1)}=\{0\}$.
Hence we have
\begin{align*}
p_{N_r^{(1)}}\bar Yp_{N_r^c}\bar Yp_{N_r^{(1)}}
=\left\vert\varphi_0\left\rangle\right\langle\varphi_0\right\vert\\
p_{N_l^{(0)}}Yp_{N_l^c}Yp_{N_l^{(0)}}
=\left\vert\varphi_1\left\rangle\right\langle\varphi_1\right\vert.
\end{align*}
Then we obtain $\delta_0=1>0$, and Assumption \ref{small2} is satisfied.
We can check for $e=-2b$ case in the same way.

%%%%%%%%%%%%%%%%%%%%%%%%%%%%%%%%%%%%
\section{Tensor Product of Linear Operators}
About tensor product of linear operators, we have the following Theorem.
It can be proved using spectral theorem given in \cite{RS}.
\begin{thm}
Let ${\cal H}_i (i=1,2)$ be separable Hilbert spaces,
and let $L_i$ be self-adjoint operators on ${\cal H}_{i}$.
Let $L$ be the self-adjoint operator on ${\cal H}_1\otimes{\cal H}_2$,
defined by
\[ L\equiv L_1\otimes 1+1\otimes L_2.\] 
Suppose that $L_2$ has a unique eigenvector $\Omega$.
Then every eigenvector of $L$ is of the form
\[ \varphi\otimes \Omega, \quad \varphi\in {\cal H}_1.\]
\label{eigentensor}
\end{thm}
\begin{align*}
\end{align*}
%%%%%%%%%%%%%%%%%%%%%%%%%%%%%%%%%%%%%%%%%%%%%%%%%%%%%%%%

\end{document}